\documentclass[11pt]{article}
\usepackage[margin=1in]{geometry}
\usepackage{setspace}
\usepackage{amsmath, amsthm, amssymb, mathtools}
\usepackage{amsfonts, bm}

\usepackage{graphicx}
\usepackage{subcaption}
\usepackage{wrapfig}
\usepackage{rotating}
\usepackage{epstopdf}

\usepackage{booktabs}
\usepackage{multirow}
\usepackage{array}

\usepackage[ruled,vlined]{algorithm2e}

\usepackage[round]{natbib}

\usepackage[english]{babel}
\usepackage{blindtext}
\usepackage{indentfirst}
\usepackage[T1]{fontenc}
\usepackage[utf8]{inputenc}
\usepackage{url}
\usepackage{microtype}
\usepackage{enumitem}
\usepackage[dvipsnames]{xcolor}
\usepackage[hidelinks]{hyperref}
\usepackage{cleveref}
\usepackage{bbm}

\numberwithin{equation}{section}
\usepackage{authblk}

\DeclareMathOperator*{\argmin}{arg\,min}
\DeclareMathOperator*{\argmax}{arg\,max}

\theoremstyle{definition} 
\newtheorem{example}{Example}[section]

\usepackage{tikz}
\usetikzlibrary{shapes, arrows.meta, positioning}

\title{Balancing Efficiency and Equity in Classroom Assignment under Endogenous Peer Effects}

\author[1]{Lei Bill Wang}
\author[2]{Zhenbang Jiao}
\author[3]{Om Prakash Bedant}
\author[1]{Haoran Wang}

\affil[1]{Department of Economics, Ohio State University}
\affil[2]{Department of Statistics, Ohio State University}
\affil[3]{Department of Electrical and Computer Engineering, Ohio State University}

\begin{document}
\maketitle
\thispagestyle{empty}
\begin{abstract}
This paper presents a three-step empirical framework for optimizing classroom assignments under endogenous peer effects, using data from the China Education Panel Survey (CEPS). 

\textbf{Step 1: Modeling Friendship Networks.}  We design \textit{PeerNN}, a neural network that mimics endogenous network formation as a discrete choice model, generating a friendship-intensity matrix ($\Omega$) that captures student popularity. 

\textbf{Step 2: Estimating Peer Effects.} We measure the peer effect friends' average 6th-grade class rank weighted by $\Omega$ on 8th-grade cognitive test score. Incorporating $\Omega$ into the linear-in-means model induces endogeneity. Using quasi-random classroom assignments, we instrument friends' average 6th-grade class rank with the average classmates' 6th-grade class rank (unweighted by $\Omega$). Our main regression result shows that a 10\% improvement in friends' 6th-grade class rank raises 8th-grade cognitive test scores by 0.13 SD. Positive $\beta$ implies maximizing (minimizing) the popularity of high (low) achievers optimizes outcomes.

\textbf{Step 3: Simulating Policy Trade-offs.} We use estimates from Step 1 and Step 2 to simulate optimal classroom assignments. We first implement a genetic algorithm (GA) to maximize average peer effect and observe a 1.9\% improvement. However, serious inequity issues arise: low-achieving students are hurt the most in the pursuit of the higher average peer effect. We propose an \textit{Algorithmically Fair GA} (AFGA), achieving a 1.2\% gain while ensuring more equitable educational outcomes.

These results underscore that efficiency-focused classroom assignment policies can exacerbate inequality. We recommend incorporating fairness considerations when designing classroom assignment policies that account for endogenous spillovers.
\end{abstract}

\newpage
\pagenumbering{arabic} 

\section{Introduction}

This paper proposes a three-step empirical framework for classroom assignment optimization in the presence of endogenous network formation and peer effect \citep{goulas2024optimal}. Imagine a middle school has an incoming cohort of students who have just finished elementary school. The middle school principal needs to assign these students to classrooms. The principal has the following two beliefs and forms an assignment strategy based on these beliefs. 

The first belief is endogenous network formation. Once the students are assigned to classrooms, they form friendship networks in their classrooms and receive more peer influence from their close friends. As a result, popular students exert more peer effects on the entire classroom.
The second belief is heterogeneous peer quality, which leads to differential peer effects. The peer effect received by a student can be positive or negative depending on whether the friends of the said student are \textit{good students} or \textit{disruptive peers}. As a result of these two beliefs, the principal wants to strategically devise a classroom assignment policy that maximizes good students' popularity and minimizes disruptive peers' popularity, ultimately \textit{maximizing the average peer effect}.



\begin{figure}[h]
    \centering
    \includegraphics[width=.95\linewidth]{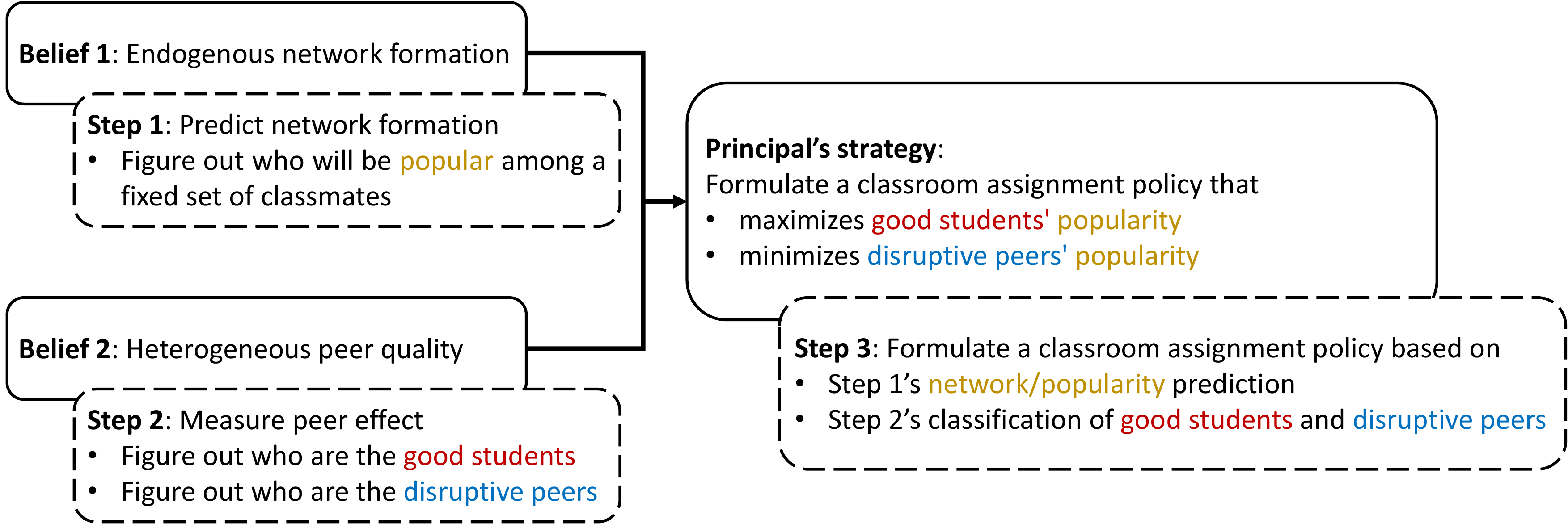}
    \caption{Beliefs and strategy of the principal (solid) and their corresponding steps (dashed).}
    \label{fig S1S2S3}
\end{figure}

Correspondingly, this paper breaks the principal's classroom assignment problem into three parts to output a classroom assignment policy that aligns with the beliefs and the principal's strategy. \Cref{fig S1S2S3} summarizes the beliefs and strategy of the principal and their corresponding steps in our three-step empirical framework.

To the best of our knowledge, our framework is the first empirical attempt for classroom assignment optimization under \textit{both} endogenous network formation and peer effect. Given the novelty of the problem, we make some adaptations of current methods pulled in from different literature. Next, we briefly introduce each step's general idea, related literature, and result summary.

\subsection*{Step 1: Modelling friendship formation}
We predict middle school classroom friendship formation with students' 6th-grade information as predictors. The goal of Step 1 is to figure out which students will be popular given a fixed set of classmates, as written in \Cref{fig S1S2S3}. 

By combining two streams of network literature, we propose a new estimation method, named PeerNN, for linkage intensity. The first stream of literature \textbf{predicts linkage using neural network}.
Mainstream link prediction studies use neural networks \citep{ahmed2016efficient, wu2016link, wang2014friendbook, dhelim2022survey, chunaev2020community, su2022comprehensive, samanta2021measure, ling2023deep}. Though both our work and these cited papers use neural networks to approximate complicated social network formation processes, we intentionally avoid techniques such as aggregating neighbors' information used in the cited papers. We exclusively use students' individual information as predictors. This approach is necessitated by the absence of linkage information under counterfactual classroom assignments. The second literature \textbf{models friendship preferences using a micro-founded discrete choice} \citep{graham2017econometric,mele2022structural}. We incorporate the microfoundation concepts, such as discrete choice, homophily, and transitivity, into PeerNN's architecture.

An advantage of PeerNN over many other network models is that it does not need expensive linkage data as a response variable. Instead, we can use easier-to-collect aggregate relational data\citep{mccormick2015latent,breza2020using,mccormick2020network}. ARD also imposes less of a threat to individual identification and other privacy-related problems.

\textbf{Preview of Step 1 results}: PeerNN captures social network characteristics such as gender homophily, outperforming the uniform friendship formation process implicitly assumed in traditional linear-in-means models widely used in peer effect literature \citep{guryan2009peer, carrell2010externalities, carrell2013natural, burke2013classroom}.

\subsection*{Step 2: Measuring peer effect}
We measure how friends' average 6th-grade class rank affects students' 8th-grade cognitive test scores. If the peer effect is positive, then students who did well in 6th-grade will be considered good students, and those who did poorly in 6th-grade will be considered disruptive peers.

We replace the restrictive uniform spillover assumption of the linear-in-means model with a more reasonable \textbf{friendship-weighted} specification. This specification leverages the nonuniform friendship formation predicted by PeerNN, addressing the limitations of the linear-in-means model, which assumes uniform spillover effects and ignores friendship dynamics \citep{bramoulle2009identification, goldsmith2013social}. The uniform spillover assumption is especially problematic for our datasets, in which the classroom size is relatively large \citep{hong2022coworker}. The average classroom size is more than 40 students.

Introducing a network into pure contextual peer effect measurement is challenging because network formation is endogenous. Unobserved student characteristics can simultaneously affect friend choice and educational outcome. We exploit the \textit{random classroom assignment} conditional on school choice to construct an instrument to conduct causal inference. We instrument friends' average 6th-grade class rank (weighted by $\Omega$) with the classmates' average 6th-grade class rank (unweighted by $\Omega$). This instrument satisfies the relevance constraint because the assigned classmates are the pool of friends that a student can pick from, so friends' average and classmates' average 6th-grade class rank correlate. The exclusion constraint is satisfied in theory because the classmates' assignments are random after conditioning on school-fixed effects, assuming the absence of correlated peer effect mechanisms that do not operate through friendship. Such an assumption is likely to be violated, e.g., competitiveness of the classroom environment, quality of classroom discussion, and teachers' adjustments for the level of instruction can correlate with the classroom assignment. To address these correlated mechanisms, we replace the classroom-random effect with classroom-level covariates to close the backdoor path. Furthermore, we test out the heterogeneous peer effect design with our dataset.  


\textbf{Preview of Step 2 results}:  All specifications' conclusions are consistent: having friends who were high-achieving students in elementary school causally increases students' 8th-grade cognitive test scores. In particular, the our baseline specification shows that a 10\% improvement in friends' 6th-grade class rank raises 8th-grade cognitive test scores by 0.13 SD. 

Step 2 result means that students who did well in elementary schools are the \textit{good students} whereas students who did poorly in elementary schools are the \textit{disruptive peers}. This slightly differs from Figure \ref{fig S1S2S3} in the sense that instead of a binary classification, we have a spectrum of good students and disruptive peers since class ranks are converted into a continuous measure of class quantile.

\subsection*{Step 3: Optimizing classroom assignment} 
In the third step, we optimize classroom assignments in terms of maximizing \textit{average} peer effects by maximizing/minimizing the popularity of the good students/disruptive peers, respectively. 

Unlike existing literature, which implicitly assumes straightforward network formation that depends on only policy but not individuals' characteristics \citep{bhattacharya2009inferring,carrell2013natural}, our approach allows dependence between friendship intensity and characteristics of all classmates in one network, making the network dependent on \textit{both} the assignment policy and the predetermined student characteristics (i.e., endogenous). 
Consequently, we cannot recast the optimization problem as a linear programming problem as done by the cited works. 
We employ a heuristic optimization method, genetic algorithm (GA), to navigate this complex problem. 

\textbf{Preview of Step 3 results}: GA outputs a classroom assignment policy that, at median, improves peer effect by 1.9\% in our counterfactual simulation. However, the output policy is extremely inequitable. In pursuing the highest average peer effect for the whole classroom, GA gives up on a few disruptive peers by predicting that they form an exclusive clique.
Consequently, students in that clique experience severely negative peer effects, but overall, the entire classroom has a higher average peer effect. 
A naive optimization of classroom assignment can lead to extremely inequitable educational outcomes. We propose a modified fairness-aware fitness function for GA and name it algorithmically fair genetic algorithm (AFGA) (the naming follows \cite{kleinberg2018algorithmic,mitchell2021algorithmic}). The new fitness function penalizes the variance of predicted peer effects within and across classrooms. AFGA outputs slightly less efficient (the improvement is 1.2\%) but much more equitable policies. 

\subsubsection*{Contribution of the paper}
Our contributions are two-fold. First, we propose a novel way of estimating friendship formation parameters as a discrete choice model. The estimation method, named PeerNN, combines the flexibility of modern machine learning algorithms and the microfoundation of economic modeling. We apply PeerNN to CEPS data and show that it can capture prominent features in the network. 
Readers can use PeerNN as an exploratory tool to look for hidden network structures when aggregated relational data is available, or it can be used as a preliminary step for peer effect estimation, as in this paper. Second, though executing our framework in real life is challenging, our attempt to optimize classroom assignment highlights the critical need to incorporate fairness considerations when optimizing the efficiency of classroom assignment policies. In pursuit of optimal average outcome, students with weaker initial academic standing are hurt the most, causing inequity concerns. We recommend explicitly placing equity safeguards when designing optimal policy under an endogenous spillover setup, and propose an intuitive way to do so.

\subsection*{Paper organization}
The rest of the paper is organized as follows: Section \ref{sec data description and data notation} describes the CEPS data and introduces the notations that we use in this paper, Section \ref{model setup} sets up the models and objective functions in our three-step framework, Section \ref{empirical results} demonstrates and explains the results that we obtain from the empirical study and Section \ref{Conclusion} concludes the paper.

\section{Notation and data description} \label{sec data description and data notation}
We use the publicly available China Education Panel Survey (CEPS) data provided by the National Survey Research Center at Renmin University of China (NSRC).
CEPS conducted surveys on 7,649 students, along with their family members and teachers, across 179 classes. For classes that are selected to participate, \textbf{all} students were surveyed.

\subsection{Classroom size and school size} 
We use $N$ to denote the number of students in a \textbf{single} classroom. If we want to emphasize classroom $c$ (or a classroom $c$ from school from $s$),  we use $N_c$ (or $N_{cs}$) to denote the class size. Additionally, we use $N_s$ to denote the number of students in school $s$. 

\subsection{Predetermined characteristics}
CEPS collected information from students and their family members regarding the students' 6th-grade details, for example: class rank (in quantile), academic subject interest, extracurricular hobbies, etc.  

When assigning students to classrooms upon school entry, the principal only has access to their information predetermined before the students enter middle school. Therefore, we exclusively use these predetermined variables (e.g. gender and locality), collectively denoted as $X$, to predict friendship formation. Occasionally, we use the notation $X_c$ (or $X_s$) for the predetermined characteristics of all students from a classroom $c$ (or from a school $s$).

One particular variable in $X$ that we repeatedly use in different steps of our methodology is students' 6th-grade class quantile, denoted as $z$. If we need to emphasize the 6th-grade quantile of all students from classroom $c$ (or of a particular student $i$ from classroom $c$ and school $s$ or of all students from school $s$), we use $z_{c}$ (or $z_{ics}$ or $z_s$) to denote that information.

\subsection{Aggregated relational data} \label{sec ARD}
Students were asked to provide aggregated relational data (ARD) of their friends. The linkage data, though surveyed in section C20 as depicted in Figure \ref{friendship survey questions} 
in Appendix \ref{Appendix Friendship questionnaire}, 
is redacted from CEPS data. Instead, ARD of the friendship network (See section C21 from Figure \ref{friendship survey questions}) is provided. Therefore, we have information on `out of the five best friends of a student, how many of them \underline{\hspace{3cm}}?' where \underline{\hspace{3cm}} can be filled by any question from section C21. 

We use ARD, denoted as $A_f$, as the response variable for PeerNN to uncover friendship formation patterns. $A_f$ has size $(N \times 10)$ and records each student's responses to 10 distinct questions about their friends' gender, locality, relationship status, etc. Each question only has three possible answers: `none', `one or two', or `most of them'. For example, one student may report none of his friends are female, most of his friends are local, and one or two of his friends are in a relationship, etc.

Apart from $A_f$, each student's own information regarding these 10 questions is also available in CEPS. The answer to each question is binary (yes or no).  For example, one student is female, local, and in a relationship, etc. We denote students' own information as $A_s$, which has the same dimension as $A_f$. 
We use $A_{fiq}$ to denote ARD information provided by student $i$ about question $q$ and $A_{sq}$ to denote all students' own characteristics for question $q$.

\subsection{Train and test sets} \label{sec train and test}
Upon entering their respective secondary schools, 5860 (from 139 classrooms) out of 7649 students (from 179 classrooms) were \textbf{randomly} assigned to classrooms, as indicated by \textit{a survey question } in CEPS. This natural experiment setup helps generate exogenous variation in data. Exploiting this data-generating property, we construct instrumental variables for causal inference analysis. To keep our input data consistent for all three steps, we use these 5860 students' data (1) as the training set for PeerNN, (2) to estimate peer effect, and (3) to design classroom assignments. The rest of the 1789 students who are non-randomly assigned to classrooms are used as test data for PeerNN.

\subsection{Data summary}
Table \ref{Summary} presents students’ demographic and background characteristics in the full, training, and test datasets. In the full sample, the average class size is approximately 45 students, and the average student age is 167.57 months (around 14 years). 47.2\% of the students are female, 50.4\% hold rural Hukou, and 8.8\% belong to non-Han ethnic groups. The average 6th-grade academic quantile is 66\%. The average father's and mother's education level is 4, corresponding to a technical secondary school or technical school degree—above junior high school but below senior high school.

In the training dataset, the average class size is 46 students, with the average age of 167.57 months (14 years). Among these students, 48\% are female, 48.5\% have rural Hukou, and 8.7\% are non-Han. Parental education levels remain consistent with those in the full sample. In the test dataset, the average age remains the same, but the gender and background composition differs slightly compared to the training. Female students make up a lower proportion (44.7\%), while the proportion with rural Hukou is higher at 56.5\%, and 9.3\% are non-Han. The test data also show a higher average class size (47 students) and slightly lower average education levels for both parents compared to the training dataset.

\begin{table}[htbp]\centering
\def\sym#1{\ifmmode^{#1}\else\(^{#1}\)\fi}
\begin{tabular}{l*{6}{c}}
\toprule
            &\multicolumn{2}{c}{Full}&\multicolumn{2}{c}{Train}&\multicolumn{2}{c}{Test}\\
            &\multicolumn{1}{c}{Mean} &\multicolumn{1}{c}{S.D.} &\multicolumn{1}{c}{Mean} &\multicolumn{1}{c}{S.D.} &\multicolumn{1}{c}{Mean} &\multicolumn{1}{c}{S.D.}\\
\midrule
Age                & 167.568 &  7.939   & 167.565 & 7.790  & 167.575 & 8.410 \\
                   &&&&&& \\
Gender             & 0.472   & 0.499    & 0.480   & 0.500  & 0.447   & 0.497 \\
&&&&&& \\
Hukou              &  0.504  & 0.500    & 0.485   & 0.500  & 0.565   & 0.496 \\
&&&&&& \\
6th-grade quantile &  0.660  &  0.233   & 0.660   & 0.236  & 0.662   & 0.223 \\
&&&&&& \\
Class size         &  45.057 &  11.529  & 45.682  & 11.869 & 47.285  & 10.243 \\
&&&&&& \\
Father's education & 4.329   &  2.024   & 4.419   & 2.061  & 4.035   & 1.871 \\
&&&&&& \\
Mother's education &  4.067  &  2.007   & 4.169   & 2.038  & 3.733   & 1.867 \\
&&&&&& \\
Ethnic nationality &  0.088  &  0.283   & 0.087   & 0.281  & 0.093   & 0.290 \\
&&&&&& \\
Number of observations & \multicolumn{2}{c}{7,649} & \multicolumn{2}{c}{5,860} & \multicolumn{2}{c}{1,789}\\
\bottomrule
\end{tabular}
\caption{Summary statistics of the full dataset, train set and test set}
\label{Summary}
\end{table}

\subsection{Test for random assignment}
The training sample is selected based on a survey question that asks the teachers whether the classes that they teach are randomly assigned. On top of that, we further conduct likelihood ratio tests to check whether class assignment is based on students’ academic rank and other characteristics. The intuition behind this approach is that if students are randomly assigned to classes within each school, then their individual characteristics should be independent of class assignment. In other words, the proportion of students with specific characteristics in each class should not differ systematically from the corresponding proportions at the school level.

To test this, we conduct a likelihood ratio test (LRT) based on the distribution of students’ specific characteristics. Let $N_c$ and $N_s$ denote the number of students in the class and school respectively and $n_c$ and $n_s$ denote the number of students with a specific characteristic in the class and school respectively. When we randomly draw a class in each school, the likelihood ratio is 
\[
\text{LR}_c = \frac{\mathcal{L}_0}{\mathcal{L}_1}=\frac{\left( \frac{N_c}{N_s} \right)^{n_c} \left(1 - \frac{N_c}{N_s} \right)^{n_s - n_c}}{\left( \frac{n_c}{n_s} \right)^{n_c} \left(1 - \frac{n_c}{n_s} \right)^{n_s - n_c}},
\]
where $\mathcal{L}_0$ is the likelihood under the null hypothesis, where is proportion in each class should equal to the school-level proportion, and $\mathcal{L}_1$ is the likelihood under the alternative hypothesis.

\begin{figure}[htb]
     \centering
     \includegraphics[width = 1\textwidth]{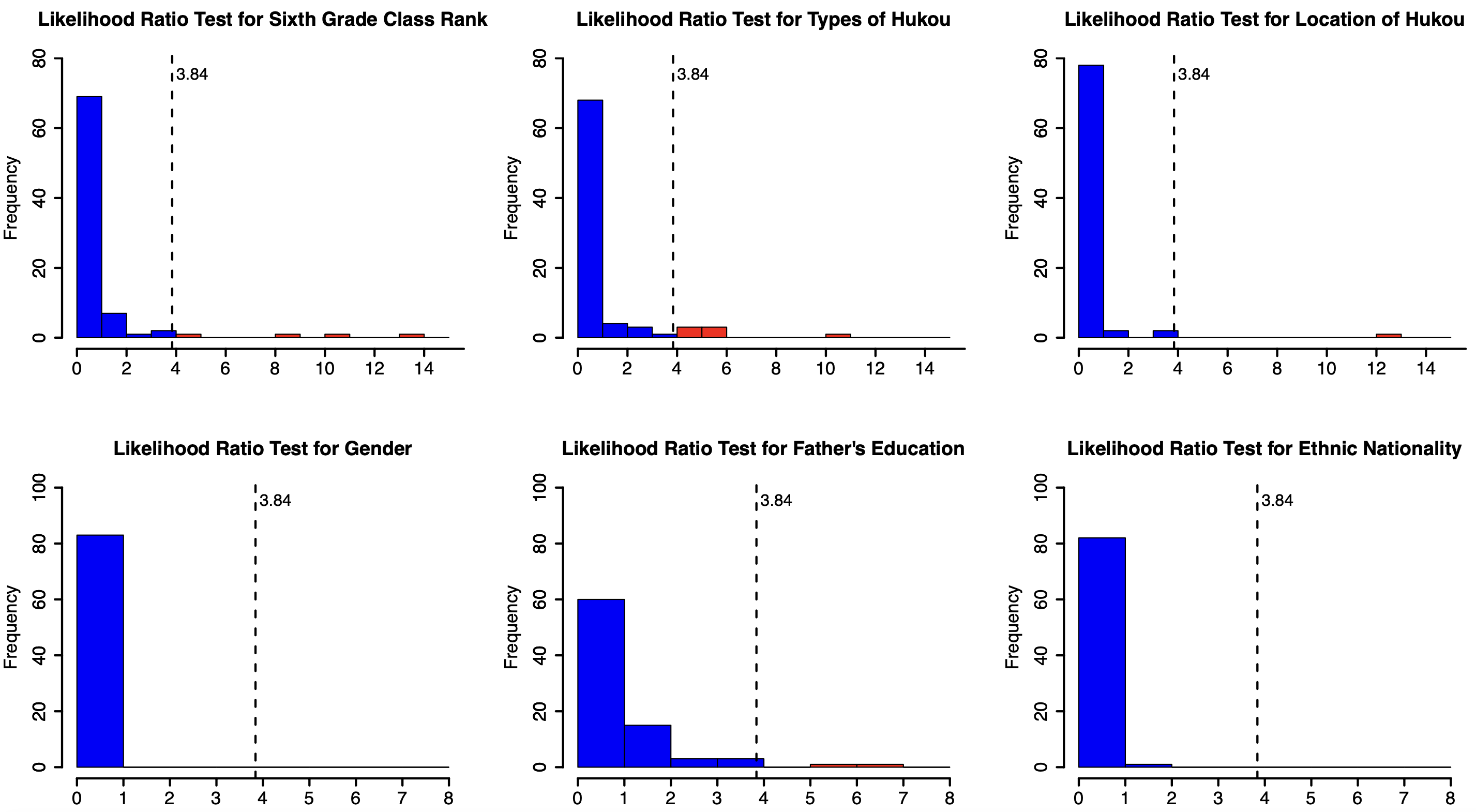}
     \caption{Likelihood ratio tests}
     \label{likelihood_ratio}
 \end{figure}

Then, we use the likelihood ratio to compute the $\chi^2$ statistics by using $\chi^2_c = -2 \log(\text{LR}_c)$ and get the $\chi^2$ distribution. We perform this test using several student characteristics: 6th-grade class academic quantile, Hukou type (rural or not), Hukou location (local or not), gender, father's education, and mother's education. Figure \ref{likelihood_ratio} displays the test results, with the black vertical line indicating the critical value. Most of the $\chi^2_c$ statistics fall below the critical threshold, suggesting that the within-class proportions of these characteristics do not significantly differ from the school-level proportions. 

Furthermore, we conduct Pearson's $\chi^2$ tests within schools for each variable, following \cite{ammermueller2009peer} and \cite{wang2021social}. The results are presented in Figure \ref{person_chi} in the Appendix. We also conduct more conventional balance tests in Appendix \ref{Appendix balance test}. The results support that the classrooms in the training set are randomly assigned. 

\section{Model setup} \label{model setup}

There are $N_s$ incoming students from middle school $s$ indexed by $i \in \mathcal{I} :=\{1,2,\dots,N_s\}$. The principal needs to assign the $N_s$ students to $|C|$ classrooms indexed by $\{1,2,\dots,|C|\}$. We denote the set of indices for classrooms as $C := \{1,2,\dots,|C|\}$. A classroom assignment policy, $\pi$, is a partition of $\mathcal{I}$ with partition cardinality $|C|$ like in \citet{goulas2024optimal}. Each subset in $\pi$ (the indices of students assigned to one classroom) is denoted as $\pi_c$ where $c \in C$, meaning that $\pi = \{\pi_c\}_{c \in C}$. The principal knows how many students should be assigned to each classroom, hence, the cardinality of $\pi_c$ is fixed before the students are assigned to classroom $c$ and denoted as $N_c$.
The ultimate goal of the principal is to devise a policy $\pi$ that maximizes peer effect. The objective function of the principal of middle school $s$ is
\begin{equation} \label{objective function}
    \max_{\pi \in \Pi} \frac{1}{N_s} \sum_{c \in C} \mathbf{1_{N_c}^{\top}} \Omega_c(\pi,X_s) z_c(\pi,z_s) \beta \quad \text{where} \quad \Omega_c(\pi,X_s) = f(X_c(\pi,X_s))
\end{equation}
where $\Pi$ is the set of all policies satisfying the cardinality constraints on both the number of classrooms $|C|$ and the size of each classroom $\{N_c\}_{c \in C}$. 
$\mathbf{1_{N_c}}$ denotes a vector of ones of length $N_c$. 

$X_c$ (an $N_c \times p$ matrix) and $z_c$ (an $N_c$ column vector) are predetermined characteristics and the 6th-grade class quantile of all students in classroom $c$. Our model allows $z_c$ to be one of the columns of $X_c$ (or not). We formally define $X_c := \{X_{s,i}\}_{i \in \pi_c}$ and $z_c := \{z_{s,i}\}_{i \in \pi_c}$, where $X_s$ and $z_s$ do not depend on $\pi$ and have $N_s$ rows, $\pi_c$ depends on $\pi$ and determines which rows of $X_s$ and $z_s$ should be included in $X_c$ and $z_c$, respectively.

\textbf{Step 1 overview} $f: \mathbbm{R}^{N_c \times p} \to [0,1]^{N_c \times N_c}$ maps predetermined student characteristics $X_c$ to a $N_c \times N_c$ friendship intensity matrix of classroom $c$, $\Omega_c$. $f$ is interpreted as the friendship formation process.  
In Section \ref{sec PeerNN}, we build a neural network architecture, PeerNN, to learn $f$. PeerNN takes $X_c$ as input and predicts $\Omega_c$ (i.e., $f(X_c)$) as an intermediate output. The loss function of PeerNN is based on the match between $\Omega_c$ and ARD provided by the students from classroom $c$.

\textbf{Step 2 overview} $\beta$ is the peer effect parameter in the regression equation we specify later in Section \ref{sec instrumental variable peer effect measurement} with the following interpretation: if a student's friends' average 6th-grade class quantile improves by 10\%, then the student's cognitive ability test score improves by $\frac{\beta}{10}$ point, holding everything else fixed. We will relax the homogeneous $\beta$ assumption in Section \ref{sec heterogeneity analysis}.

\begin{figure}[htb]
        \centering
        \includegraphics[width = \textwidth]{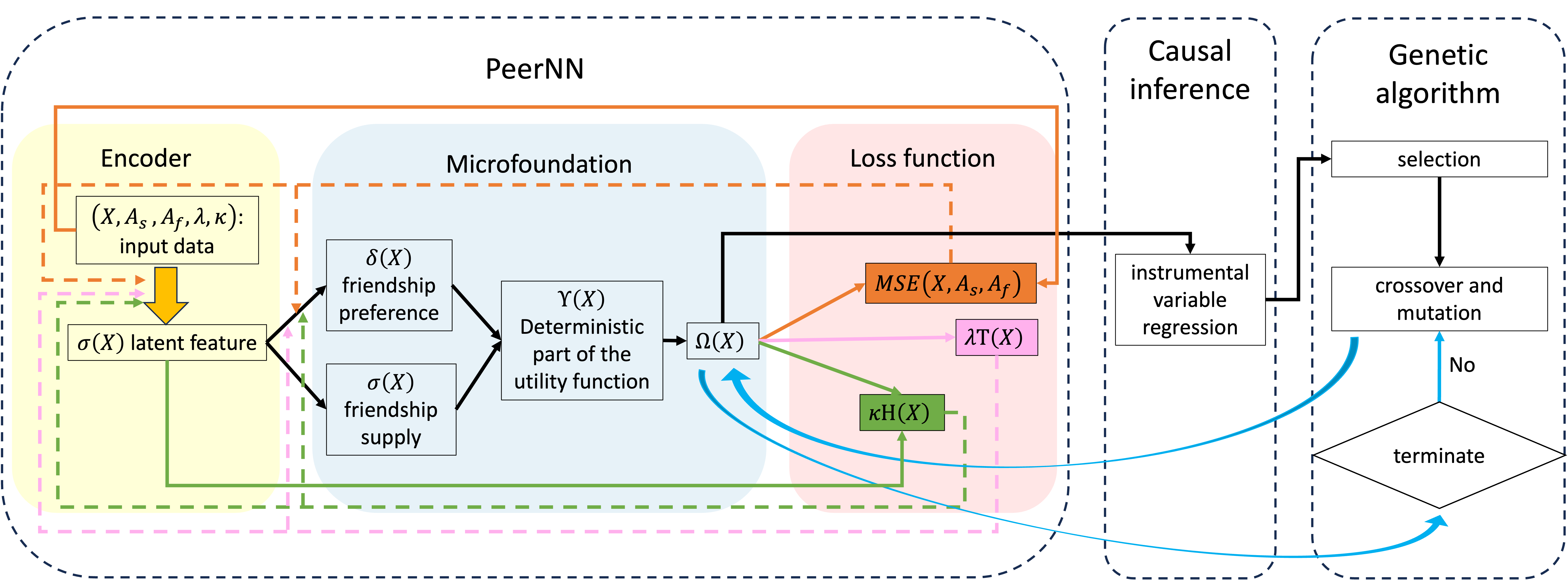}
        \caption{The three-step framework that takes in predetermined characteristics of students and outputs the optimal $\pi$. All subscripts are dropped because we illustrate the framework with respect to one \textbf{single} classroom. The first step (PeerNN) takes in $X$ and predicts $\Omega$ as an intermediate output; the second step incorporates $\Omega$ into the regression equation and estimates $\beta$; the sign of $\beta$ determines the third step (GA)'s fitness function, GA iteratively plugs in PeerNN to search for the optimal $\pi$.} 
        \label{Overall}
    \end{figure}

\textbf{Step 1-3 overview} Solving (\ref{objective function}) requires nonconventional approaches for two reasons: first, $\Pi$ is discrete, making the optimization combinatorial and NP-hard; second, $\Omega_c$ is a function of \textit{both} $\pi$ and $X_s$, making it impossible to write it as function of \textit{only} $\pi$ or recast (\ref{Overall}) as a one-step feasible programming problem like in \citet{bhattacharya2009inferring,carrell2013natural}. As shown by Figure \ref{Overall}, we solve (\ref{objective function}) with the following three steps: (i) fitting/predicting network $\Omega_c$ with PeerNN (ii) estimating peer effect parameter $\beta$ with an instrumental variable, and (iii) heuristically searching for the optimal $\pi$, that maximizes (\ref{objective function}) with GA.

\subsection{Friendship formation neural network} \label{sec PeerNN}
We build an interpretable neural network architecture, termed PeerNN, to predict friendship formation with ARD. PeerNN can be divided into four stages: (1) friendship market, (2) utility function, (3) normalization, and (4) loss function. We provide concrete examples for each stage in Appendix \ref{Appendix examples section}. The description pertains to a \textbf{single} classroom. Since the classroom is fixed, we drop the subscript $c$ in Section \ref{sec PeerNN} for a cleaner notation. 




\subsubsection*{Stage 0: Latent feature encoder} We encode our raw data $X$ to a lower dimensional embedding $\sigma(X)$ with a fully connected hidden layer that contains ten nodes, $\sigma(X)$ is an $(N \times 10)$ matrix. 
Stage 0 serves as a feature extraction step for Stages 1-4 and is estimated as part of PeerNN.

\subsubsection*{Stage 1: Friendship market}
In friendship market, students supply themselves as their latent features, $\sigma(X)$. 
Students' preference parameters $\delta$ (dimension $(N \times 10)$) are functions of $\sigma(X)$. The ith row of $\delta$ matrix, $\delta_i$ is the preference parameters of student $i$. $\delta_{ik}$ indicates how much student $i$ prefers to have a friend with a higher value of latent characteristic $k$.
We use two fully connected layers to model preference parameters $\delta(X)$:
\begin{equation*}
    \delta(X) = {\rm ReLu}({\rm ReLu}(\sigma(X) W^{(1)})W^{(2)}) \quad \text{where} \quad \sigma(X) = {\rm ReLu}(XW^{(0)}).
\end{equation*}
Example \ref{example matching} illustrates the concept of preference parameters. We strongly encourage readers who are not familiar with machine learning/neural networks literature to refer to Appendix \ref{Appendix examples section} for the intuition behind PeerNN.

\subsubsection*{Stage 2: Deterministic part of utility function}
We obtain the outer product of $\delta(X)$ and $\sigma(X)$ and denote the outer product as $\Upsilon$, an $N \times N$ matrix, where $\Upsilon_{ij} = \sum_{k = 1}^K \delta_{ik}\sigma_{jk} = (\delta \cdot \sigma')_{ij}$. $\Upsilon_{ij}$ corresponds to a concept in the structural applied microeconomics literature \citep{mcfadden_conditional_1974, train2009discrete}: the deterministic part of the utility function for student $i$ picking student $j$ as best friend. 

\subsubsection*{Stage 3: Normalization}
We manually set $\Upsilon_{ii}$ as negative infinity since students cannot make friends with themselves.  Utility for student $i$ to consider student $j$ as a best friend is subject to a utility shock $\xi_{ij} \overset{iid}{\sim}$ Gumbel distribution: $U_{ij} = \Upsilon_{ij} + \xi_{ij}$. Student $i$ picks his best friend based on who gives him the highest utility, $\argmax_{j} U_{ij}$. 
Knowing the joint distribution of $\{\xi_{ij'}\}_{j' = 1}^N$, the probability of student $i$ picking student $j$ as best friend is given by softmax function 
\[
P(\argmax_{j^{'}} U_{ij^{'}} = j) = \frac{\exp{(\Upsilon_{ij})}}{\sum_{j'}^{N} \exp{(\Upsilon_{ij'})}} := \Omega_{ij}.
\]
$\Omega$ is obtained by applying the softmax function on $\Upsilon$ row-wise. This formulation interprets $\Omega_i$, ith-row of the friendship intensity matrix, as the probability student $i$ considers each classmate as a best friend. Example \ref{example Stage 2 and 3} in Appendix~\ref{Appendix examples section} demonstrates how Stage 2 and 3 work.

\subsubsection*{Stage 4: Loss function} PeerNN's loss function is a weighted sum of three components: (1) fitted value's MSE, (2) homophily penalty, and (3) transitivity penalty. 

\textbf{Fitted value's MSE} 
The dependent/response variable of PeerNN is the ARD that students report about their friends $A_f$. 
 Let $B_i$ denote the number of friends student $i$ reported. We can randomly draw $B_i$ friends for $i$th student based on $\Omega_i$, summarize his friends' information about question $q$ with $A_{sq}$, and predict his response about their friends for question $q$, $A_{fiq}$. Our prediction for $A_{fiq}$ is denoted as $\hat{A}_{fiq}(\Omega_i,A_{sq}, B_i)$. The MSE component of the loss function is 
 $E[(\hat{A}_{fiq} - A_{fiq})^2]$. The expectation is taken over all students and questions.

Two issues arise. Due to CEPS's survey design explained in Section \ref{sec ARD}, an exact mapping from $(\Omega_i, A_{sq}, B_i)$ to $A_{fiq}$ requires nonlinear operations, rendering MSE to have no closed form. Moreover, sampling friends without replacement based on the probability vector $\Omega_i$ is intractable in backpropagation. See appendix \ref{Appendix MSE} for more details on these issues. To address the two problems, 
we make two compromises: (1) approximating nonlinear operations with multiple linear functions, and (2) drawing friends with replacement. 
We can decompose MSE into ${\rm Bias}^2$ and ${\rm Var}$. With the two compromises, we derive closed-form expressions for ${\rm Bias}^2$ and ${\rm Var}$, exponentially reducing the computational complexities and making the backpropagation process tractable.
The description of the two compromises and the derivations of the closed-form expressions are shown in Appendix \ref{Appendix MSE}. Here we only present the closed-form expressions:
\begin{align*}
    {{\rm Bias}}^2 =& \frac{1}{NQ}\sum_{i=1}^{N} (a_{B_i}\Omega_i A_s + b_{B_i} - A_{fi})^{\top}(a_{B_i}\Omega_i A_s + b_{B_i} - A_{fi})  \\
    {{\rm Var}} =& \frac{1}{NQ}tr(\sum_{i=1}^{N} a_{B_i}^2 A_s^{\top} {\rm Var}_{\Omega_i} A_s)
\end{align*}
where $(j,k)$ entry of $\rm{Var}_{\Omega_i}= B_i \Omega_{ij} (1 - \Omega_{ij})$ when $j = k$ and $(j,k)$ entry of $\rm{Var}_{\Omega_i} = - B_i \Omega_{ij} \Omega_{ik}$ when $j \neq k$.
$Q$ denotes the number of ARD questions and $(a_{B_i},b_{B_i})$ are as follows. The explanation for these numerical values of $(a_{B_i},b_{B_i})$ is in Appendix \ref{Appendix loss function}. Here, we tabulate the values of $(a_{B_i},b_{B_i})$.
\begin{table}[h]
    \centering
    \begin{tabular}{c c c c c c}
    \toprule
         $B_i$ & 1 & 2 & 3 & 4 & 5 \\
         \midrule
         $b_{B_i}$&1. & 1.090 &1.154 &1.2 & 1.333\\
         $a_{B_i}$&1.5 &  0.727 & 0.654 & 0.5  &0.4\\
         \bottomrule
    \end{tabular}
\end{table}


\textbf{Homophily penalty} It is well established in social network formation literature that people tend to make friends who share similarities with themselves. This is called the homophily effect. To incorporate this social network phenomenon into PeerNN's loss function, we compute the average friend's profile and penalize the difference between this profile and the students' own latent characteristics $H(X) = \lVert \Omega(X)\sigma(X) - \sigma(X) \rVert_2 ^2$.

 Instead of incentivizing PeerNN to predict friends who are similar in raw data $X$, we penalize the difference between average friends' profile $\Omega(X)\sigma(X)$ and students' own latent features $\sigma(X)$. The choice between raw data $X$ and latent feature $\sigma(X)$ is not of importance. However, different columns of $X$ do not have the same distributions, for example, gender is a Bernoulli variable whereas class quantile is uniformly distributed. Penalizing directly using $X$ may require us to add a few more weights into the training parameters. Using latent features introduces more flexibility to the homophily penalty and alleviates the need for more weight parameters. 

\textbf{Transitivity penalty} By incorporating transitivity into PeerNN, we incentivize the model to form a clustering pattern in predicting friendship. $\Omega_{ij}$ measures the probability of a pair of students $(i,j)$ forming friendship, $(\Omega\Omega)_{ij}$ measures how likely a third student $k$ is considered as best friend by student $i$ \textbf{and} considers student $j$ as best friend. If students $i$ and $j$ are very likely to be best friends, then they are very likely to have a `mutual' best friend in a third student $k$. 
While $\Omega_{ij}$ tells us the probability of student $i$ picking student $j$ as best friends, $\Omega\Omega_{ij}$ informs the probability that student $i$ considers student $k$ as best friend who considers student $j$ as best friend. If transitivity is prevalent in classroom friendship, then we should observe that when $\Omega_{ij}$ is large then $\Omega\Omega_{ij}$ is also large. However, since we enforce the diagonal of $\Omega$ to be zero, we impose the same constraint on $\Omega\Omega$ and normalize each row to be a multinomial distribution. In short, the transitivity loss is computed as $T(X) = \lVert \Omega - (I - {\rm diag}(\Omega\Omega))^{-1} (\Omega\Omega - {\rm diag}(\Omega\Omega)) \rVert_2^2$.

The loss function is a weighted sum of the three components: ${{\rm MSE}} + \kappa H(X) + \lambda T(X) = {{\rm Bias}}^2 + {{\rm Var}} + \kappa H(X) + \lambda T(X)$ as shown in Figure \ref{Overall}. So far, we have described the loss function for one classroom. The loss is summed over all classrooms.

\textbf{Additional notes on loss function:}
When running the empirical analysis, we find it helpful to downweigh variance. The benefit of downweighing variance can be intuitively rationalized by considering network density. We defer the explanation to Appendix \ref{Appendix downweighing variance} as it does not alter the main idea of why the loss function makes sense. The modified loss function can be written as ${{\rm Bias}}^2 + \mu{{\rm Var}} + \kappa H(X) + \lambda T(X)$, where $(\mu, \kappa, \lambda)$ are hyperparameters. 
We can avoid cross-validation for tuning hyperparameters by examining whether $\Omega$ satisfies commonly known properties of social networks like gender homophily, presence of central nodes, and clustering. Such examination of $\Omega$'s interpretability can help pin down a range of reasonable values of $(\mu, \kappa, \lambda)$. We substantiate our claim in Appendix \ref{Appendix tuning parameter selection}.

\subsection{Instrumental variable peer effect measurement} \label{sec instrumental variable peer effect measurement}

\subsubsection{Incorporating network information into linear-in-means}
The existing peer effect literature largely adopts a linear-in-means specification:

\begin{equation}
    \underbrace{y_{ics}}_{\text{Student i's 8th-grade cognitive test score}} = \beta \tilde{W}_{ics} + X_{ics}\gamma + \underbrace{\theta_s}_{\text{Fixed Effect (FE)}} + \underbrace{\mu_{cs}}_{\text{Random Effect (RE)}}  + \epsilon_{ics} \label{linear in means specification}
\end{equation}
where $y_{ics}$ is cognitive ability test score of student $i$ from class $c$ school $s$, $\tilde{W}_{ics} = \frac{\sum_{j=1,j\neq i}^{N_{cs}} z_{jcs}}{N_{cs}-1}$ is student $i$'s middle school classmates' average 6th-grade class quantile, 
$X_{ics}$ is a set of control variables for student $i$ such as the student's 6th-grade class quantile, gender, race, and parents' education level. $\theta_s$ and $\mu_{cs}$ are the school fixed effect and classroom random effect, respectively. 
We can rewrite linear-in-means model as the $y_{cs} = \beta W_{cs} z_{cs} + X_{cs} \gamma + \theta_s +\mu_{cs} + \epsilon_{cs}$ where

\begin{equation*}
    W_{cs} = \begin{pmatrix}
        0 &\frac{1}{N_{cs}-1} &\frac{1}{N_{cs}-1} &\dots & \dots &\frac{1}{N_{cs}-1} \\
        \frac{1}{N_{cs}-1} &0 &\frac{1}{N_{cs}-1} &\dots &\dots &\frac{1}{N_{cs}-1} \\
        \vdots & \vdots &\vdots & \vdots &\vdots &\vdots\\
        \frac{1}{N_{cs}-1} &\frac{1}{N_{cs}-1} &\frac{1}{N_{cs}-1} &\dots & \dots & 0
    \end{pmatrix}.
\end{equation*}
Under our framework, the linear-in-means model implicitly assumes that all students exert the same amount of peer influence on all their classmates as evidenced by the $W_{cs}$ matrix which has column sums equal to 1 for each column. This rules out the possibility of heterogeneous influence among all classmates and hence, no matter how we formulate classroom assignments, all students have the same total influence on all their classmates (sum up to 1) and there is no strategic scope to optimize classroom assignments. Our specification (\ref{second stage equation}) deviates from this setup.

Taking the $\Omega$ modeled by PeerNN as the true friendship intensity matrix, we replace the linear-in-means specification with a friendship-weighted specification (\ref{second stage equation}), generating a more meaningful interpretation of $\beta$ for policymaker. The $\beta$ in (\ref{linear in means specification}) measures peer effect from an average \textbf{classmate}, whereas $\beta$ in (\ref{second stage equation}) measures the peer effect from an average \textbf{friend}. (See Appendix \ref{Appendix linear in means} for detail of linear-in-means model and its comparison to (\ref{second stage equation})).
\begin{equation}
    y_{ics} = \beta \tilde{\Omega}_{ics}  + X_{ics}\gamma + \theta_s + \mu_{cs}  + \epsilon_{ics} \label{second stage equation}
\end{equation}
where $\tilde{\Omega}_{ics} = {\sum_{j=1}^{N_{cs}}\Omega_{ijcs} z_{jcs}}$ is student $i$'s friends' average 6th-grade ranking. 

\subsubsection{Addressing endogeneity with IV}\label{Addressing endogeneity with IV}
\begin{figure}[htb]
    \centering
    \includegraphics[width=0.65\linewidth]{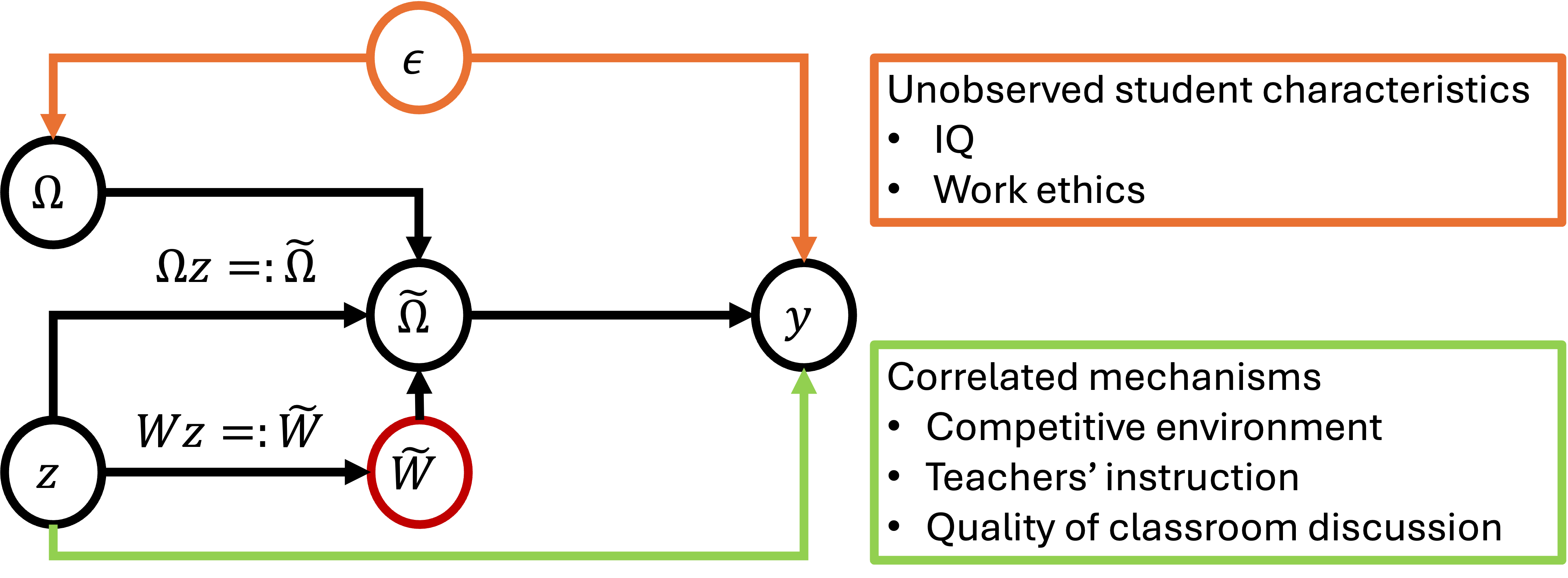}
    \caption{Directed acyclic graph for the illustration of causality. The orange components depict the endogeneity problem, which we address in Section \ref{Addressing endogeneity with IV}. The green components are the potentially correlated mechanisms which are addressed in Section \ref{sec correlated mechanisms}.}
    \label{fig IV DAG}
\end{figure}

We face an endogeneity issue with the $\Omega$ matrix, which we illustrate with example \ref{endogeneity example} and depict with the orange components in Figure \ref{fig IV DAG}. 
Student $i$'s friendship selection, ${\Omega}_{ics}$, is correlated with some unobserved characteristics of student $i$ in $\epsilon_{ics}$. $\epsilon_{ics}$ can also directly impact student $i$'s cognitive test score. We resort to the classical instrumental variable (IV) approach to address the endogeneity. The instrument that we use is the average 6th-grade class quantile of middle school classmates, $\tilde{W}$ in (\ref{linear in means specification}). A valid instrument needs to fulfill two properties: exclusion constraint and relevance constraint. 

The \textbf{relevance constraint} requires the instrument and the endogenous variable to be correlated. Both average 6th-grade class rank ($W_{cs}z_{cs}$) and friends' weighted average 6th-grade class rank ($\Omega_{cs}z_{cs}$) are dependent on all classmates' 6th-grade class rank ($z_{cs}$). Therefore, the relevance constraint is satisfied. In Figure \ref{fig IV DAG}, the relevance constraint is represented by the arrow pointing from $\tilde{W}$ to $\tilde{\Omega}$. 

The \textbf{exclusion constraint} requires the instrument (average 6th-grade class rank for all classmates) to be uncorrelated with the unobserved confounder. Barring the correlated mechanisms shown as the green component in Figure \ref{fig IV DAG}, this requirement is satisfied in theory since all classroom assignments are random conditional on school choice, which is a covariate included in (\ref{second stage equation}). Hence, any confounder that is individual $i$-specific and predetermined before middle school should not correlate with the average characteristics of the classmates that are randomly assigned to him. We will address the correlated mechanisms (green components) separately in Section \ref{sec correlated mechanisms}.

We estimate (\ref{second stage equation}) using a two-stage method that resembles the classical two-stage least-squares procedure. First, run OLS for the reduced form equation (\ref{first stage equation}) and get fitted values of $\tilde{\Omega}$.
    \begin{equation}
        \label{first stage equation}
        \tilde{\Omega}_{ics} = \pi_1 \underbrace{\frac{\sum_{j=1,j\neq i}^{N_{cs}} z_{jcs}}{N_{cs}-1}}_{\text{Classmates' average 6th-grade ranking}} + X_{ics}\delta + \phi_s + \eta_{ics}
    \end{equation}
Second, substitute $\tilde{\Omega}$ in (\ref{second stage equation}) with the fitted value and run MLE assuming normal $\mu$ and $\epsilon$. We also implement other standard methods for IV and random effect \citep{balestra1987full,baltagi1981simultaneous,han2016efficiency} to check the robustness of the regression results.

\subsubsection{Addressing correlated mechanisms} \label{sec correlated mechanisms}
While our paper focuses on the peer effect that operates through friendships $\Omega$ (and hence, $\tilde{\Omega}$), there are other correlated mechanisms that do not operate through friendship, e.g., competitiveness of the classroom environment, quality of classroom discussion, and teachers' adjustments for the level of instruction. In these cases, peer effect can exist between two students even when they are not friends, i.e., $z$ has a direct impact on $y$ without going through $\tilde{\Omega}$. Consequently, all classmates' 6th-grade class rank $z$ becomes a confounder, as shown by Figure \ref{fig IV DAG}. To address this concern, we augment the base specifications with classroom-level covariates to close the backdoor path depicted in green. These control variables include the information about the 6th-grade class ranks of the entire classroom including the student themselves (25th, 75th quantile of $z_c$ and whether $z_{ics}$ is above the median of $z_c$). The rationale behind this augmentation is that the correlated mechanisms are likely functions of these additional control variables. For example, how competitive a classroom is likely a function of top students' performance (25th quantile $z_c$), whereas teachers' instruction pace is more likely a function of the bottom students' performance (75th quantile of $z_c$). 

Including other functions of $z_c$, e.g., quantiles of $z_c$, into the specifications potentially leads to collinearity problems, which may make our estimate not significant when there is a true peer effect. Hence, we conduct further robustness checks by replacing the quantile of $z_c$ with the quantile of 6th-grade persistence measures. The persistence measures close the backdoor path similarly to the quantiles of $z_c$: classroom competitiveness depends on how persistent students are, and teacher adjust their teaching pace based on how persistent students are. 

\subsubsection{Policy significance}
The two $\beta$s in (\ref{linear in means specification}) and (\ref{second stage equation}) have different interpretations. For linear-in-means model, $\beta$ in (\ref{linear in means specification}) is the marginal effect of middle school \textbf{classmates' average} 6th-grade class quantile on the outcome; for friendship-weighted specification, $\beta$ in (\ref{second stage equation}) is the marginal effect of middle school \textbf{friends' weighted average} 6th-grade class quantile on the outcome.\footnote{Both marginal effect parameters (the two $\beta$s) reflect causality: the former comes from a randomized experiment whereas the latter uses IV to establish causality.} The former has very little policy implication to classroom assignment since the linear-in-means model essentially restricts how much total peer effect each student can exert on other classmates (See detail in Appendix \ref{Appendix linear in means}). Our friendship-weighted specification relaxes this restriction: a popular student is allowed to exert more influence on all his classmates than another less popular student. 
This allows us to devise an average-maximizing class assignment if $\beta$ in (\ref{second stage equation}) is estimated to be statistically significant. 

\subsection{Genetic algorithm (GA) that optimizes class assignment}
Since most of the schools have exactly two classrooms surveyed in CEPS, 
we assume the principal would like to optimize peer effect in these two classrooms, $c = \{1, 2\}.$\footnote{Note that even though having more classrooms makes running GA more time-consuming, it makes the improvement of GA policy over random assignment more salient. This is because as we add in more classrooms, there are greater possibilities/configurations of classroom assignments for us to pull positively influencing student groups into one classroom and separate ``disruptive peers'' from those who are more likely to be influenced by them.}


\subsubsection{Genetic algorithm (GA)}

In this section, we illustrate some specific features of the GA design that we use to optimize classroom assignment policy.
We elaborate on the GA design that we use with Algorithm \ref{GA algorithm}, which is presented in Appendix \ref{Appendix GA algorithm}. We summarize GA's essence for those interested in the general concept of GA as follows: 
\begin{itemize}
    \item Step 0: Initialize a random classroom assignment that satisfies the gender ratio constraint. 
    \item Step 1: Draw a Bernoulli variable, if it is 1, then randomly pick a pair of students, one from each classroom, and swap them, evaluate the new classroom assignment's fitness like in Step 0, repeat step 1. Else if the Bernoulli variable is 0, proceed to step 2.
    \item Step 2: GA randomly swaps a pair of students to generate a new classroom assignment and evaluates its fitness.
    \item Step 3: GA repeats Step 2 with a new pair of students swapped each time. The number of repetitions is user-specified.
    \item Step 4: GA selects the best classroom assignment among those evaluated in Steps 1-3 and updates the optimal classroom assignment as that.
    \item Step 5: Repeat Steps 1-4 with the updated optimal classroom assignment as the initial assignment until some stopping criterion is satisfied.

\end{itemize} 
Note that this is a schematic summary of GA and many modifications are made when we apply GA. Readers should refer to Algorithm \ref{GA algorithm} for the exact implementation of our GA (and AFGA) design.

We use a selection-mutation-crossover-termination four-step design. One special feature of our GA design is the incorporation of gender-ratio constraint: 
a swap that violates this gender ratio constraint will not be permitted. 
Given that all schools in our sample have close to a 50\% gender ratio, multitudes of candidate policies that satisfy the gender ratio requirement are guaranteed to exist. Additionally, we constrain that the classrooms' sizes differ by at most 1, ensuring the classrooms are of comparable sizes. See Appendix \ref{Appendix GA algorithm} for a detailed description of the institutional constraint and the parameter values that we set for GA.


\subsubsection{Fitness function and fairness consideration}

Our schematic summary of GA picks the average peer effect as the fitness function used to evaluate a policy. Mathematically, the mean predicted peer effect for all students is written as $\frac{1}{\sum_{c = \{1,2\}} N_c} \sum_{c = \{1,2\}} \beta\mathbf{1_{N_c}^{\top}}\tilde{\Omega}_c$, where $\tilde{\Omega}_c = \Omega_c z_c$ is a vector of each student's friends' 6th-grade average class quantile. This fitness function specification potentially causes algorithmic fairness problems. GA may maximize the average at the expense of certain small groups of students' educational outcomes. We actively combat this problem by penalizing both the standard errors of peer effect within classrooms ${\rm SE}_{\{\beta\tilde{\Omega}_1\}}$ and ${\rm SE}_{\{\beta\tilde{\Omega}_2\}}$ as well as standard error of peer effect across classrooms ${\rm SE}_{\{\beta\tilde{\Omega}\}}$, where $\tilde{\Omega} = (\tilde{\Omega}_1, \tilde{\Omega}_2)$. 
The two fitness functions are given as follows:
\begin{align}
    {\rm fit}_{{\rm GA}}(C_1,C_2) =& \frac{1}{\sum_{c = \{1,2\}} N_c} \sum_{c = \{1,2\}} \beta\mathbf{1_{N_c}^{\top}}\tilde{\Omega}_c \label{Fit_GA}\\
    {\rm fit}_{{\rm AFGA}}(C_1,C_2) =& \frac{1}{\sum_{c = \{1,2\}} N_c} \sum_{c = \{1,2\}} \beta\mathbf{1_{N_c}^{\top}}\tilde{\Omega}_c -  \sum_{c = \{1,2\}} \phi{\rm SE}_{\{\beta\tilde{\Omega}_c\}} - \rho {\rm SE}_{\{\beta\tilde{\Omega}\}} \label{Fit_AFGA}.
\end{align}
where $C_1$ and $C_2$ denote the set of students assigned to the two classrooms, respectively. Note that $(C_1, C_2)$ fully pins down the class assignment policy, $\pi$. The penalties are weighted by $\phi$ and $\rho$. 
Throughout the paper, we set $\phi = \rho$.

\section{CEPS data fitting results} \label{empirical results}
We implement our three-step empirical framework with CEPS data. This section outlines the key findings of our application.

\subsection{Friendship formation pattern}

\begin{figure}[htb]
\centering
\subfloat[$\Omega$ heat map for classroom 8 from train set]{\label{fig:8}\includegraphics[width=.45\linewidth]{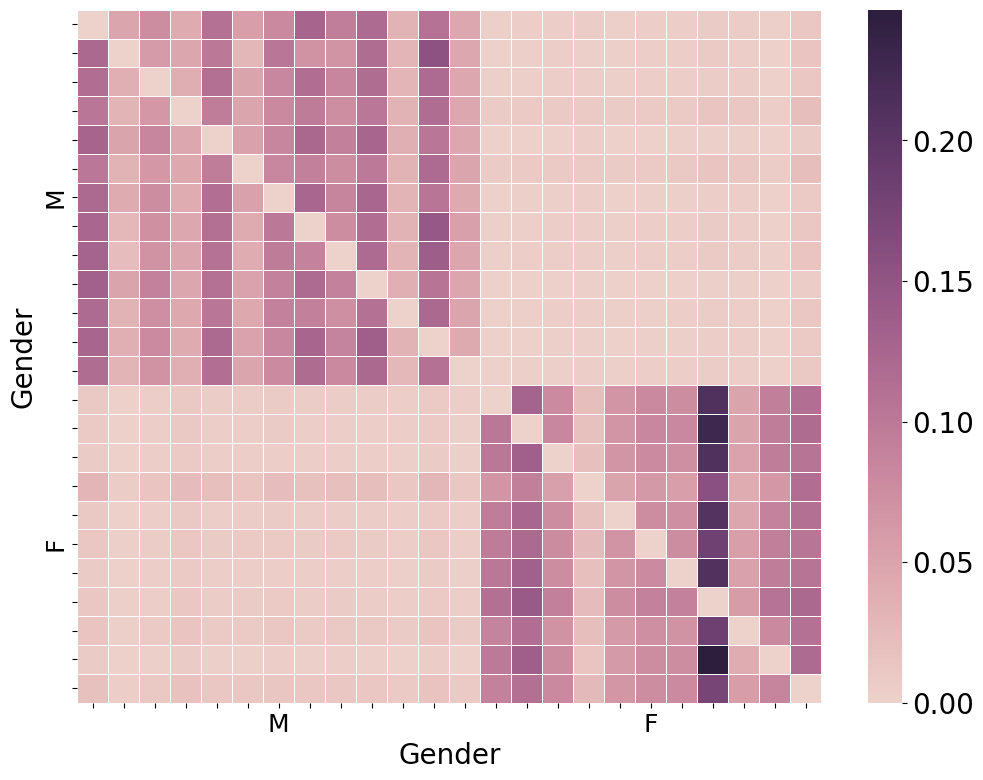}}
\hfill
\subfloat[$\Omega$ heat map for classroom 2 from test set]{\label{fig:2}\includegraphics[width=.45\linewidth]{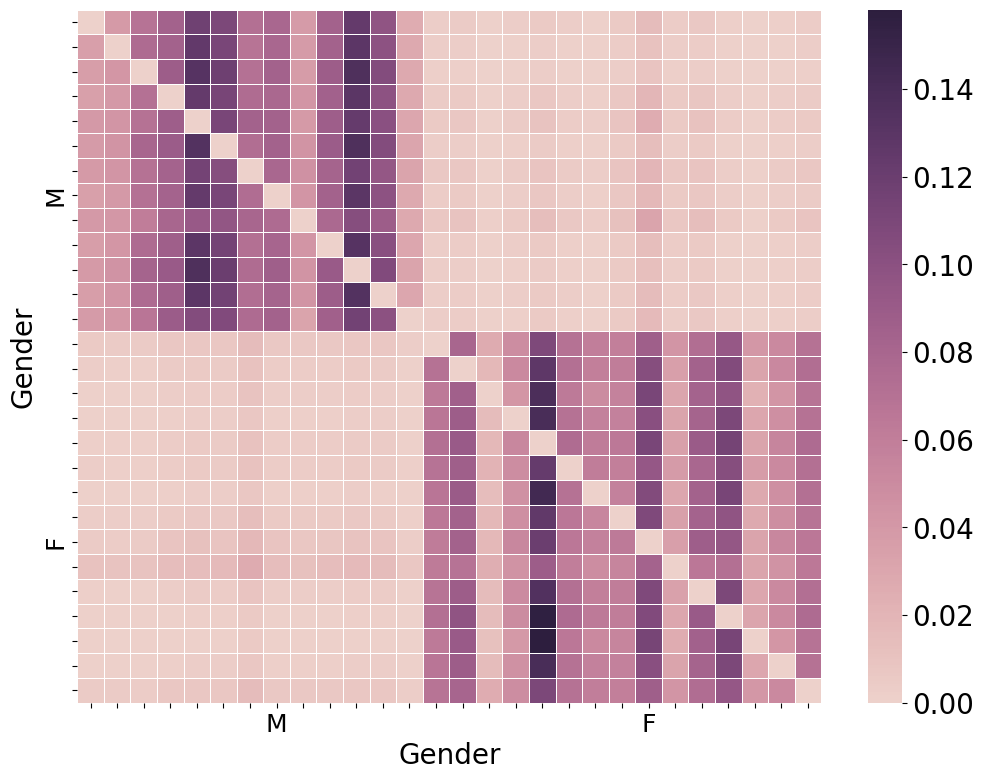}}

\caption{Heat maps for two classrooms to demonstrate that PeerNN prediction aligns with known property of social network: (1) gender homophily (2) presence of central nodes 
}
\label{heat maps FIM 8 and 2}
\end{figure}

PeerNN can capture the most salient features in the classroom network. First, it predicts extremely strong gender homophily. We reorder all students by gender in the first classroom in our dataset and plot a heat map of $\Omega$, the adjacency-probability matrix (See Figure \ref{heat maps FIM 8 and 2}). Two dark diagonal blocks indicate that students are more likely to make friends of the same gender as themselves. Second, PeerNN finds some popular individual students as indicated by the dark columns. These students are the central nodes of the classroom networks. Many social networks exhibit a centrality property where certain nodes are particularly well-connected with other nodes. In Step 3, GA will attempt to make good students the central nodes. 

\begin{figure}[htb]
    \centering
    \includegraphics[width = \textwidth]{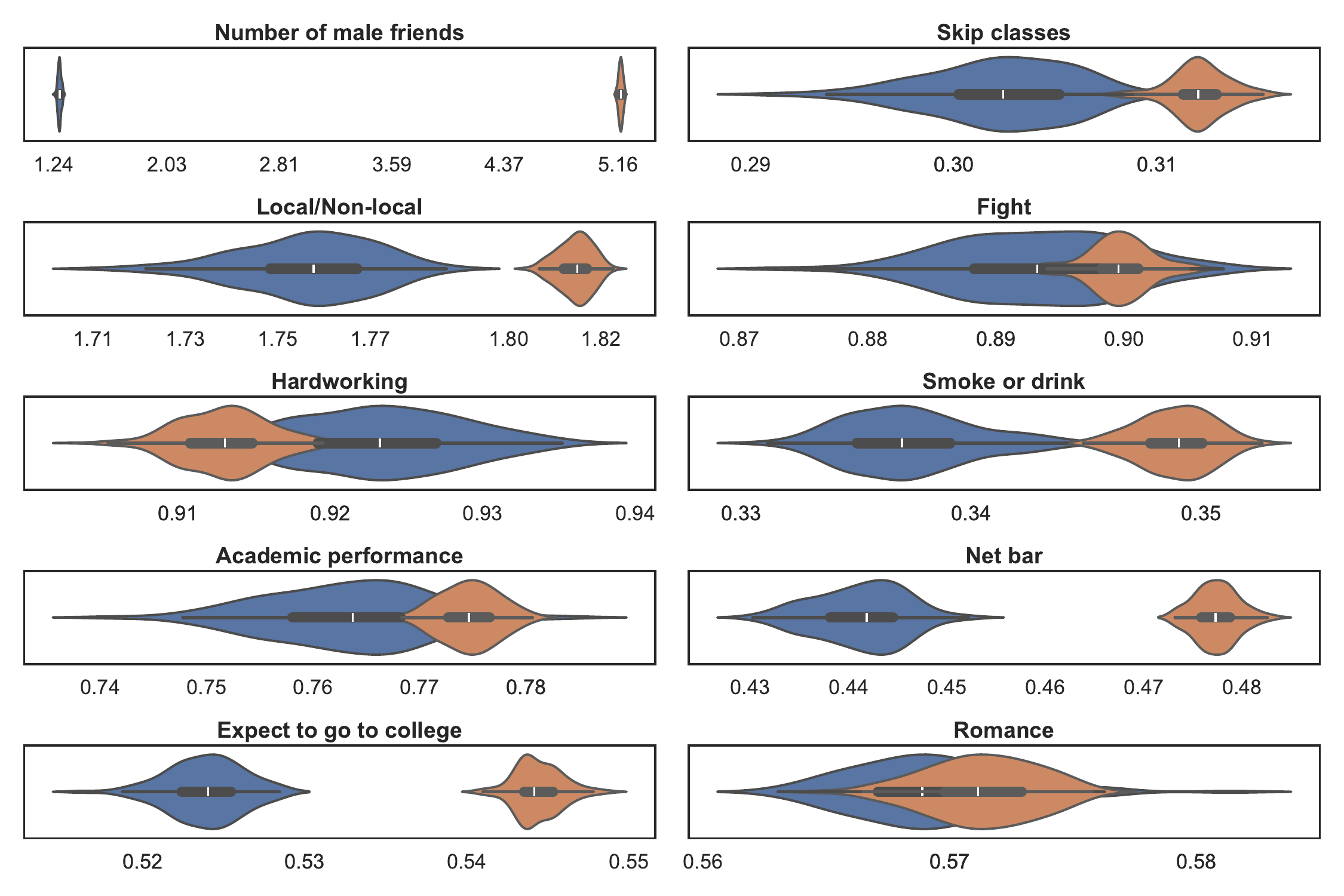}
    \caption{PeerNN prediction performance full version. Blue and orange color distributions represent the empirical densities of PeerNN and linear-in-means models' prediction errors, respectively. PeerNN outperforms the linear-in-means model in nine out of ten traits. The only trait that PeerNN performs worse than the linear-in-means model. It is likely due to how students evaluate whether they are hardworking or not may systematically differ from how they evaluate whether their friends are hardworking or not. Also, note that the scale for the `hardworking' trait is much smaller than some of the other traits such as gender and local.}
    \label{figure PeerNN prediction performance full version}
\end{figure}

We compare PeerNN with the conventional linear-in-means model. CEPS asks students to provide ARD responses about their friends in terms of ten distinct traits. We use $\Omega$ to predict friendship formation for all students and compare our prediction with the ARD responses. How we compute prediction error is detailed in Algorithm \ref{prediction error algorithm} in Appendix \ref{Appendix friendship formation pattern}.
In Figure \ref{figure PeerNN prediction performance full version}, we present violin plots of prediction error. For half of the ten traits including gender, local/nonlocal, smoke or drink, our model predicts better for all 1000 rounds. PeerNN also predicts better than the linear-in-means model for the rest of the traits except for `hardworking'. It could be due to the fact students' own evaluation of whether they are hardworking systematically differs from their friends' evaluation of them. Nevertheless, 
Overall, our model beats the linear-in-means model in terms of out-of-sample prediction by a huge margin.

\subsection{Peer effect measurement}
\subsubsection{Who are the good students/disruptive peers?}
We estimate (\ref{second stage equation}) and confirm the presence of a positive peer effect from students who did well in elementary school. We refer to students who did well in elementary school as ``good students'' and students who did not do well in elementary school as ``disruptive peers'' therein. Note that since class quantile is a continuous measure, there is a spectrum of ``good students'' and ``disruptive peers'', we use these two terms for easier reference. 

Our estimation result of (\ref{second stage equation}) shows that if student $i$'s friends' weighted average 6th-grade class quantile improves by 10\%, his 8th-grade cognitive ability test score improves by 0.1082 on a scale of 5, or equivalently, 13\% of standard deviation. To put it more concretely, the average cognitive test score is 0.3608 which lands the student at the 43rd quantile of the score distribution (the score distribution is left-skewed). An 0.1082 increase in cognitive test score improves the student's ranking from 43rd quantile to 51st quantile. In Table \ref{regression result full}, we show six specifications' results. 
One, Linear-in-means model without classroom random effect; two, Linear-in-means model with classroom random effect; three, Instrumental variable without classroom random effect; four, Instrumental variable with classroom random effect; five, Instrumental variable with classroom-level controls; six, Instrumental variable with classroom-level controls and classroom-level random effect.

\begin{table}[htbp]\centering
\def\sym#1{\ifmmode^{#1}\else\(^{#1}\)\fi}

\begin{tabular}{l*{6}{c}}
\toprule
            &\multicolumn{2}{c}{Linear-in-means} &\multicolumn{4}{c}{IV}\\
\midrule
Peer's Rank &   0.973\sym{***} &   0.980\sym{**}   & 1.075\sym{***} & 1.082\sym{**} & 0.920\sym{***} & 1.034\sym{*}  \\
 &     (0.244)         &     (0.462)    & (0.269) & (0.510) & (0.307) & (0.578) \\[1.5ex]
Own Rank & 1.019\sym{***}  & 1.022\sym{***}  & 0.995\sym{***} & 0.997\sym{***} & 0.999\sym{***} & 1.001\sym{***}  \\
& (0.038)  & (0.039)  & (0.038) & (0.038) & (0.038) & (0.038) \\[1.5ex]
Age & -0.012\sym{***}  & -0.012\sym{***}  & -0.012\sym{***} & -0.012\sym{***} & -0.012\sym{***} & -0.012\sym{***} \\
& (0.001)  &  (0.001)    & (0.001)  & (0.001) & (0.001) & (0.001) \\[1.5ex]
Sex & -0.010  & -0.010    & -0.011 & -0.010 &  -0.011 &  -0.011  \\
& (0.017)  & (0.017) & (0.017)  & (0.017) & (0.017) & (0.017) \\[1.5ex]
Father's education & 0.020\sym{***}  & 0.017\sym{***} & 0.020\sym{***} & 0.018\sym{***} & 0.020\sym{***}  & 0.018\sym{***}  \\
& (0.006)  &  (0.006)    & (0.006) & (0.006) & (0.006) & (0.006) \\[1.5ex]
Mother's education & 0.006  & 0.006  & 0.005 & 0.005 &  0.005  &  0.005  \\
& (0.006)  &  (0.006)    & (0.006)  & (0.006) & (0.006) & (0.006) \\[1.5ex]
Ethnic nationality & 0.013  & 0.022    & 0.012 & 0.021 & 0.009 & 0.019   \\
& (0.040)  &  (0.040)  & (0.040) & (0.040)  & (0.040) & (0.040) \\[1.5ex]
Teacher's sex &       &         &      &  & 0.107\sym{***} & 0.107\sym{***}     \\
&        &         &         &   & (0.034) & (0.065)      \\[1.5ex]
Teacher's working &       &         &      &  & 0.004  & 0.003    \\
experience  &        &         &         &   & (0.003)  & (0.005)     \\[1.5ex]
Proportion of  &       &         &      &  & 0.000   & 0.004   \\
female  &        &         &         &   & (0.005)    & (0.010)    \\[1.5ex]
Proportion of  &       &         &      &  & 0.002 & 0.005     \\
rural Hukou  &        &         &         &   & (0.003)   & (0.006)     \\[1.5ex]
\midrule
Number of observations  &   5,860    &  5,860   &  5,860 &  5,860    &  5,822 &  5,822  \\[1ex]
Adjusted \(R^{2}\)&   0.382    &      &   0.382  &  & 0.384 &   \\[1ex]
- Log Likelihood &     &  5,845.390    &       &  5,845.291 & & 5,820.787 \\
\midrule
Results from first stage\\[1ex]
Peer's Rank&  &      & 0.905\sym{***} & 0.905\sym{***}   & 0.909\sym{***}  & 0.909\sym{***}     \\
&    &  & (0.015) & (0.015)  &  (0.018)  &  (0.018)  \\[1ex]
Adjusted \(R^{2}\)&  &    & 0.755 &  & 0.756 & 0.756 \\
\midrule
School FE & YES & YES & YES & YES & YES & YES \\[1ex]
Class RE & NO & YES & NO & YES & NO & YES \\
\bottomrule
\multicolumn{5}{l}{\footnotesize Standard errors in parentheses}\\
\multicolumn{5}{l}{\footnotesize \sym{*} \(p<0.10\), \sym{**} \(p<0.05\), \sym{***} \(p<0.01\)}\\
\end{tabular}
\caption{The Peer Effects of Sixth Grade Rank on Cognitive Test Score}
\label{regression result full}
\end{table}

We also implement more standard methods for IV with random effect specification to estimate (\ref{second stage equation}) \citep{baltagi1981simultaneous,balestra1987full}. The estimated coefficients for peer effect are 0.975 (0.268) and 1.075 (0.270), respectively (See Table \ref{Baltagi and BVK} in Appendix for the full results). The numbers in the parentheses are standard errors. The regression results for all six specifications and all estimation methods are very close to each other, demonstrating robustness. 

The regression results 
opens up the strategic scope of classroom assignments. 
Making good students the central nodes and disruptive peers the peripheral nodes of the social network increases the entire classroom's peer effect. 

\subsubsection{Examining robustness to the correlated mechanisms}
We conduct robustness checks due to concerns arising from correlated mechanisms by controlling for variables related to class-level characteristics of classmates. Table \ref{robustness1} shows the results from regressions that include school fixed effects, as well as additional controls such as the class-level 25th and 75th percentiles of 6th-grade class rank ($z_c$), interaction terms between these percentiles and the proportion of female students, and an indicator for whether a student’s own rank is above the class median. We also include all baseline controls from the main regression, including class-level covariates. All results from the first to the last column indicate that peers' ranks have a positive effect on students' cognitive scores. However, the coefficient in the fourth column is statistically insignificant, likely due to multicollinearity between the percentile-based control variables of $z_c$ and the instrument classmates' average rank (average $z_c$), which may inflate the standard error and reduce statistical significance. 

The issue of multicollinearity becomes more pronounced when class random effects are included, as shown in columns (1) to (4) of Table \ref{robustness2}. This is expected to some extent because all the class-level control variables and classroom indicators for the random effects are multicollinear, so the regression results rely on the normal assumption imposed on the random effect parameters. Although the estimates for peer's rank are not statistically significant in these specifications, the direction remains the same, and the magnitude of the effects remains close to our baseline specifications. 

Moreover, we conduct further robustness checks by replacing the quantile of $z_c$ by other measures of persistence to alleviate the multicollinearity issue. Following \cite{golsteyn2021impact} and \cite{zou2024peer}, persistence refers to a student's ability to continue and persevere in their efforts, even when faced with difficulties. Persistence reflects a student’s attitude toward learning and can serve as a proxy for study-related ability. The CEPS includes questions on students’ persistence \textit{in their 6th grade}. Following \cite{zou2024peer}, we measure persistence using three survey items: “I would try my best to go to school even if I was not feeling very well or I had other reasons to stay at home”; “I would try my best to finish even the homework I dislike”; and “I would try my best to finish my homework even if it would take me quite a long time.” Each question is rated on a 1–5 scale (from strongly disagree to strongly agree). We sum the scores and standardize the resulting measure to have a mean of 0 and standard deviation of 1 \citep{zou2024peer}. Then, we define students with high persistence level whose level is higher than the mean of the persistence of the sample. We calculate the proportion of students with high persistence level in each class (normal mean level and leave-out mean) and add the new control in our main regression. 

The results in Table \ref{robustness3} show that peer rank has a significant positive effect on students' cognitive scores. Only the estimate in column (2) is not statistically significant, but it is marginally insignificant with a p-value of 0.107. 
Overall, our results remain robust after controlling for persistence, further supporting our baseline results that having friends who are high-achieving in elementary school increases one's cognitive test score in 8th grade.

\begin{table}[htbp]\centering
\def\sym#1{\ifmmode^{#1}\else\(^{#1}\)\fi}

\begin{tabular}{l*{5}{c}}
\toprule
            &\multicolumn{1}{c}{(1)} &\multicolumn{1}{c}{(2)} &\multicolumn{1}{c}{(3)} &\multicolumn{1}{c}{(4)} &\multicolumn{1}{c}{(5)}\\
\midrule
Peer's Rank &   0.977\sym{*} &   0.862\sym{*}   & 1.098\sym{**} & 0.684 & 0.980\sym{***} \\
 &     (0.587)     &     (0.458)    & (0.532) & (0.491) & (0.311)  \\[1.5ex]
Own Rank & 0.999\sym{***}  & 0.999\sym{***}  & 0.999\sym{***} & 0.999\sym{***} & 0.941\sym{***}  \\
& (0.038)  & (0.038)  & (0.038) & (0.038) & (0.060)  \\[1.5ex]
25 percentile  &  -0.038     &         &     &  &     \\
  &  (0.335)   &         &         &   &   \\[1.5ex]
75 percentile  &       &  0.066     &      &  &   \\
  &           &  (0.386)     &   &  &  \\[1.5ex]
25 percentile $\times$ &       &         &  -0.005    &  &   \\
Proportion of female  &        &         &   (0.012)      &   &   \\[1.5ex]
75 percentile $\times$  &       &         &  & -0.012   &   \\
Proportion of female  &        &         &         & (0.021)  &   \\[1.5ex]
$I[rank>median rank]$  &       &         &      &  &   0.034 \\
  &        &         &         &   &  (0.027) \\
\midrule
Number of observations  &  5,822    &  5,822  &  5,822 &  5,822    &  5,822  \\[1ex]
Adjusted \(R^{2}\)&   0.384    &  0.384    &   0.384  & 0.384  & 0.384   \\
\midrule
Results from first stage\\[1ex]
Peer's Rank  &  0.832\sym{***}   & 0.856\sym{***} & 0.834\sym{***} & 0.842\sym{***} & 0.906\sym{***} \\
&  (0.033) & (0.026) & (0.030)  &  (0.028) & (0.018) \\[1ex]
Adjusted \(R^{2}\)& 0.757 &  0.757  & 0.757 & 0.757 & 0.756 \\
\midrule
School FE & YES & YES & YES & YES & YES  \\[1ex]
Class RE & NO & NO & NO & NO & NO\\
\bottomrule
\multicolumn{5}{l}{\footnotesize Standard errors in parentheses}\\
\multicolumn{5}{l}{\footnotesize \sym{*} \(p<0.10\), \sym{**} \(p<0.05\), \sym{***} \(p<0.01\)}\\
\end{tabular}
\caption{Robustness Check with School Fixed Effects}
\label{robustness1}
\end{table}

\begin{table}[htbp]\centering
\def\sym#1{\ifmmode^{#1}\else\(^{#1}\)\fi}

\begin{tabular}{l*{5}{c}}
\toprule
            &\multicolumn{1}{c}{(1)} &\multicolumn{1}{c}{(2)} &\multicolumn{1}{c}{(3)} &\multicolumn{1}{c}{(4)} &\multicolumn{1}{c}{(5)}\\
\midrule
Peer's Rank &   1.325 &   1.352   & 1.273 & 1.162 & 1.092\sym{*} \\
 &     (1.068)     &     (0.869)    & (0.968) & (0.920) & (0.580)  \\[1.5ex]
Own Rank & 1.001\sym{***}  & 1.001\sym{***}  & 1.001\sym{***} & 1.001\sym{***} & 0.945\sym{***}  \\
& (0.038)  & (0.038)  & (0.038) & (0.038) & (0.059)  \\[1.5ex]
25 percentile  &  -0.202     &         &     &  &     \\
  &  (0.627)   &         &         &   &   \\[1.5ex]
75 percentile  &       &  -0.340   &      &  &   \\
  &           &  (0.695)     &   &  &  \\[1.5ex]
25 percentile $\times$ &       &         &  -0.007    &  &   \\
Proportion of female  &        &         &   (0.024)      &   &   \\[1.5ex]
75 percentile $\times$  &       &         &  & -0.007   &   \\
Proportion of female  &        &         &         & (0.039)  &   \\[1.5ex]
$I[rank>median rank]$  &       &         &      &  &   0.033 \\
  &        &         &         &   &  (0.027) \\
\midrule
Number of observations  &  5,822    &  5,822  &  5,822 &  5,822    &  5,822  \\[1ex]
-Log Likelihood &   5,820.289   &  5,820.118   & 5,823.564   & 5,823.092  & 5,822.737  \\
\midrule
Results from first stage\\[1ex]
Peer's Rank  &  0.832\sym{***}   & 0.856\sym{***} & 0.834\sym{***} & 0.842\sym{***} & 0.906\sym{***} \\
&  (0.033) & (0.026) & (0.030)  &  (0.028) & (0.018) \\[1ex]
Adjusted \(R^{2}\)& 0.757 &  0.757  & 0.757 & 0.757 & 0.756 \\
\midrule
School FE & YES & YES & YES & YES & YES  \\[1ex]
Class RE & YES & YES & YES & YES & YES\\
\bottomrule
\multicolumn{5}{l}{\footnotesize Standard errors in parentheses}\\
\multicolumn{5}{l}{\footnotesize \sym{*} \(p<0.10\), \sym{**} \(p<0.05\), \sym{***} \(p<0.01\)}\\
\end{tabular}
\caption{Robustness Check with School Fixed Effects and Class Random Effects}
\label{robustness2}
\end{table}

\begin{table}[htbp]\centering
\def\sym#1{\ifmmode^{#1}\else\(^{#1}\)\fi}

\begin{tabular}{l*{4}{c}}
\toprule
            &\multicolumn{1}{c}{(1)} &\multicolumn{1}{c}{(2)} &\multicolumn{1}{c}{(3)} &\multicolumn{1}{c}{(4)}\\
\midrule
Peer's Rank &  0.847\sym{***} &  0.913  &  0.852\sym{***} & 0.962\sym{*} \\
 &     (0.307)     &     (0.558)    & (0.307) & (0.563) \\[1.5ex]
Own Rank & 0.999\sym{***}  & 1.001\sym{***}  & 1.003\sym{***} & 1.003\sym{***}   \\
& (0.038)  & (0.038)  & (0.038) & (0.038) \\[1.5ex]
Proportion of students    & 0.648\sym{***}   &  0.586\sym{**} &     &     \\
with high persistence level  &  (0.143)   &  (0.263)    &         &     \\[1.5ex]
Proportion of students with high  &       &   &  0.565\sym{***}    &  0.326   \\
persistence level (leave one out)   &           &     & (0.140)  & (0.248)   \\

\midrule
Number of observations  &  5,822    &  5,822  &  5,822 &  5,822   \\[1ex]
Adjusted \(R^{2}\)& 0.387 &    & 0.386 &  \\[1ex]
-Log Likelihood &    &  5,818.819  &   & 5,820.449   \\
\midrule
Results from first stage\\[1ex]
Peer's Rank  &  0.907\sym{***} & 0.907\sym{***} & 0.907\sym{***} & 0.907\sym{***}  \\
&  (0.018) & (0.018) & (0.018)  &  (0.018) \\[1ex]
Adjusted \(R^{2}\)& 0.757 &  0.757  & 0.757 & 0.757  \\
\midrule
School FE & YES & YES & YES & YES   \\[1ex]
Class RE & NO & YES & NO & YES \\
\bottomrule
\multicolumn{5}{l}{\footnotesize Standard errors in parentheses}\\
\multicolumn{5}{l}{\footnotesize \sym{*} \(p<0.10\), \sym{**} \(p<0.05\), \sym{***} \(p<0.01\)}\\
\end{tabular}
\caption{Further Robustness Check with Persistence Level among Students}
\label{robustness3}
\end{table}

\subsubsection{Who are more (or less) affected by their peers?} \label{sec heterogeneity analysis}
For completeness of our study, we examine \textit{heterogeneous} peer effects of 6th-grade class rank based on gender, father's education, and Hukou type using our instrumental variable (IV) approach. Table \ref{Hetero effects} presents the partial results. See the full results in Table \ref{Hetero effects full} in the Appendix.

The first and second columns display the results by gender. Higher peers’ weighted average 6th-grade class quantile is positively associated with cognitive test scores for both males and females, though the effects are less statistically significant compared to the main regression. The effect for males is marginally significant, while the effect for females is marginally insignificant; however, the magnitudes are similar across both groups. These findings suggest that higher-achieving peers can positively influence cognitive outcomes for both genders, and there is no substantial gender difference in the strength of peer effects. For heterogeneity by father's education, we classify students into two groups: those whose fathers have more than a middle school education (relatively high education) and those whose fathers have at most a middle school education (relatively low education). The third and fourth columns show that higher peers’ 6th-grade class quantile has a significantly positive effect on the cognitive scores of students whose fathers have relatively low education, but an insignificant effect for those with relatively high-educated fathers. This suggests that students from less advantaged family backgrounds may benefit more from exposure to higher-achieving peers, and students from more advantaged backgrounds are already well-supported and less reliant on peer influence. Finally, the fifth and sixth columns report the results by Hukou type. The estimated peer effects are significantly positive for both rural and non-rural Hukou students and are similar in magnitude to the main results. This indicates that peer effects from 6th-grade class rank do not differ between students with rural and non-rural Hukou.

\begin{table}[htbp]\centering
\def\sym#1{\ifmmode^{#1}\else\(^{#1}\)\fi}
\begin{tabular}{l*{6}{c}}
\toprule
            &\multicolumn{2}{c}{Gender}&\multicolumn{2}{c}{Father's education}&\multicolumn{2}{c}{Types of Hukou}\\
            &\multicolumn{1}{c}{Male} &\multicolumn{1}{c}{Female} &\multicolumn{1}{c}{Low} &\multicolumn{1}{c}{High} &\multicolumn{1}{c}{Rural} &\multicolumn{1}{c}{Others}\\
\midrule
Peer's Rank &   1.000\sym{*}&  0.946   &   1.255\sym{***}   & 0.565 & 1.215\sym{**}   & 1.143\sym{*} \\
 &     (0.581)         &     (0.592)    & (0.582)  & (0.687) &  (0.619) &  (0.650) \\
Own Rank & 0.916\sym{***}  & 1.117\sym{***}  & 0.990\sym{***} &  1.037\sym{***}  & 0.984\sym{***}  & 1.008\sym{***} \\
& (0.054)  & (0.053)  & (0.053)  & (0.056) & (0.055)  &  (0.053)  \\
Number of observations  &   3,048    &  2,812   &  3,147     &  2,713 & 2,845  & 3,015  \\
- Log Likelihood   &  3,244.771    &  2,611.928   &   3,283.262 & 2,561.519  & 2,953.243  & 2,924.823  \\
\midrule
School FE & YES & YES & YES & YES & YES & YES \\
Class RE & YES & YES & YES & YES & YES & YES \\
\bottomrule
\multicolumn{5}{l}{\footnotesize Standard errors in parentheses}\\
\multicolumn{5}{l}{\footnotesize \sym{*} \(p<0.10\), \sym{**} \(p<0.05\), \sym{***} \(p<0.01\)}\\
\end{tabular}
\caption{Heterogeneous Peer Effects of Sixth Grade Rank on Cognitive Test Score}
\label{Hetero effects}
\end{table}

All the heterogeneous peer effect analyses are consistent with our classification of good students and disruptive peers. On average, high-achieving elementary school students become good students in middle school, influencing their friends positively; low-achieving elementary school students become disruptive peers in middle school, influencing their friends negatively.

\subsection{Optimal classroom assignment policy design} \label{discussion on class assignment}
\subsubsection{Overall improvement with uncertainty quantification}
This section shows that GA/AFGA improves the efficiency of the classroom assignment policy. We examine Step 3 with a fixed set of PeerNN parameters with the following steps.
For a school, (1a) Randomly assign students to classrooms (reject the assignment if institutional constraint is violated and redraw until no violation), call it R1, compute average peer effect for R1. (1b) Randomly assign students to classrooms (reject the assignment if institutional constraint is violated and redraw until no violation), call it R2, apply GA to R2, call the output policy R2+ compute average peer effect for R2+. (1c) Repeat (1b) two more times and call the output policy R3+ and R4+. Compute the average peer effect for R3+ and R4+. (1d) Repeat step (1b) and (1c) with AFGA with the weight on equity set as 0.5, 1 and 1.5. Repeat steps (1a-1d) for all schools. For each school, we simulate 13 policies in total, 1 random assignment (R1), 4 of our methods times 3 repetitions (R2+,R3+,R4+) for each method.

\Cref{table improvement summary} reports our methods' improvement in average peer effect in percentage over a random assignment policy. Here are the 5 quantile (we report one-sided results because we only care about whether our policy performs better than random assignment), median, and mean of the improvement.

\begin{table}[htbp]
\centering
\begin{tabular}{lccc}
\toprule
Method & 5\% Quantile & Median & Mean \\
\midrule
GA & 0.2413 & 1.9129 & 2.0625 \\
AFGA 0.5 & 0.0477 & 1.8207 & 1.9358 \\
AFGA 1 & 0.0290 & 1.7471 & 1.8242 \\
AFGA 1.5 & 0.0164 & 1.1570 & 1.3175 \\
\bottomrule
\end{tabular}
\caption{Performance Comparison of Methods}
\label{table improvement summary}
\end{table}

\begin{figure}[htbp]
    \centering
    \includegraphics[width=.65\linewidth]{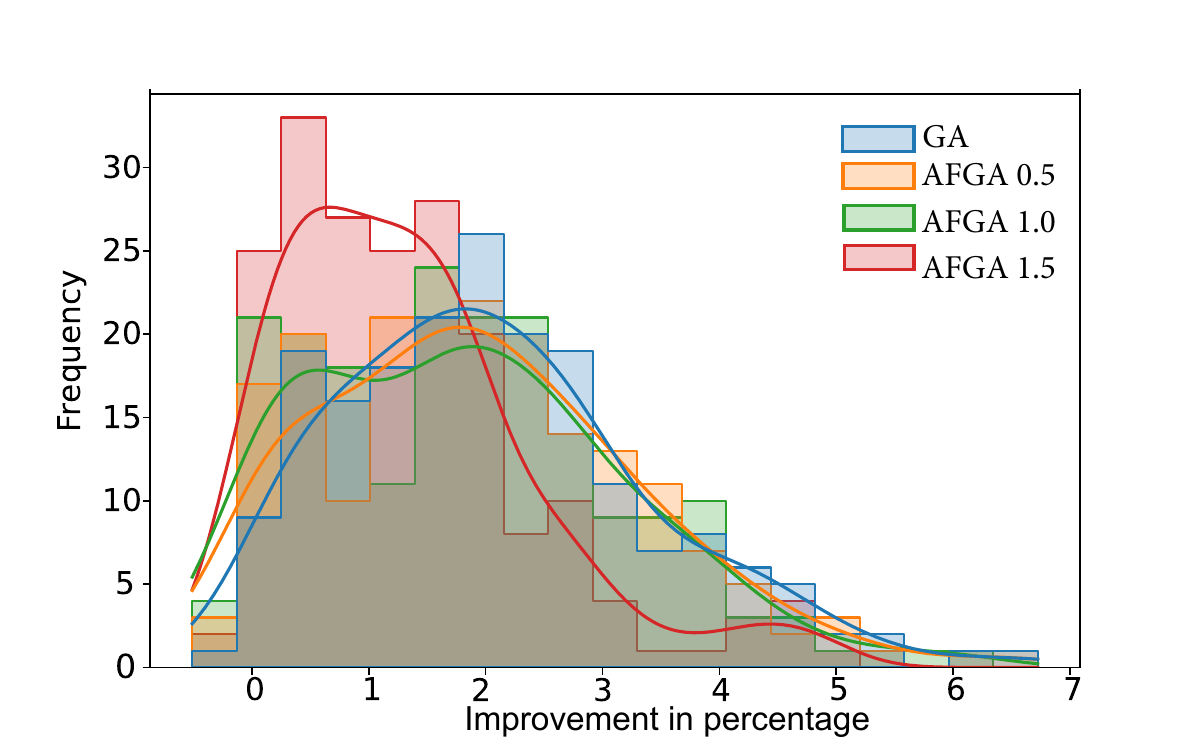}
    \caption{The empirical distribution of the improvement demonstrates that the improvement is largely positive. As the weight put on fairness increases, the empirical distribution shifts left.}
    \label{fig improvement_summary}
\end{figure}

The improvement is positive and statistically significant at 5\% level. At the median level, the improvement is 1.9\% for GA and 1.2\% for AFGA when $\phi = \rho =1.5$. This result shows that the heuristic optimization method, GA, makes non-trivial modifications to the initial classroom assignment policies. Moreover, the decrease in improvement down the rows makes sense because as we increase the weight, we are sacrificing efficiency for greater equity. See \Cref{fig improvement_summary} for a better visualization of the improvement's statistical significance and trade-off between efficiency and equity.

\subsubsection{Inequity problem}
Although there is a statistically significant improvement in the average peer effect $\tilde{\Omega}$, we encounter an inequity problem when implementing GA. In contrast, AFGA outputs policies that are more efficient than random assignment \textit{and} more equitable than GA policies.

We find it clearest to explain the problem with two classrooms from one school as an example. In this section, we use classrooms 5 and 6 from CEPS as the running example. We supplement many more examples in Appendix \ref{Appendix class assignment results} to show that the extreme inequity problem exists across the entire sample, not just the example that we include in the main text.

\begin{figure}
    \centering
    \includegraphics[width =.8\textwidth]{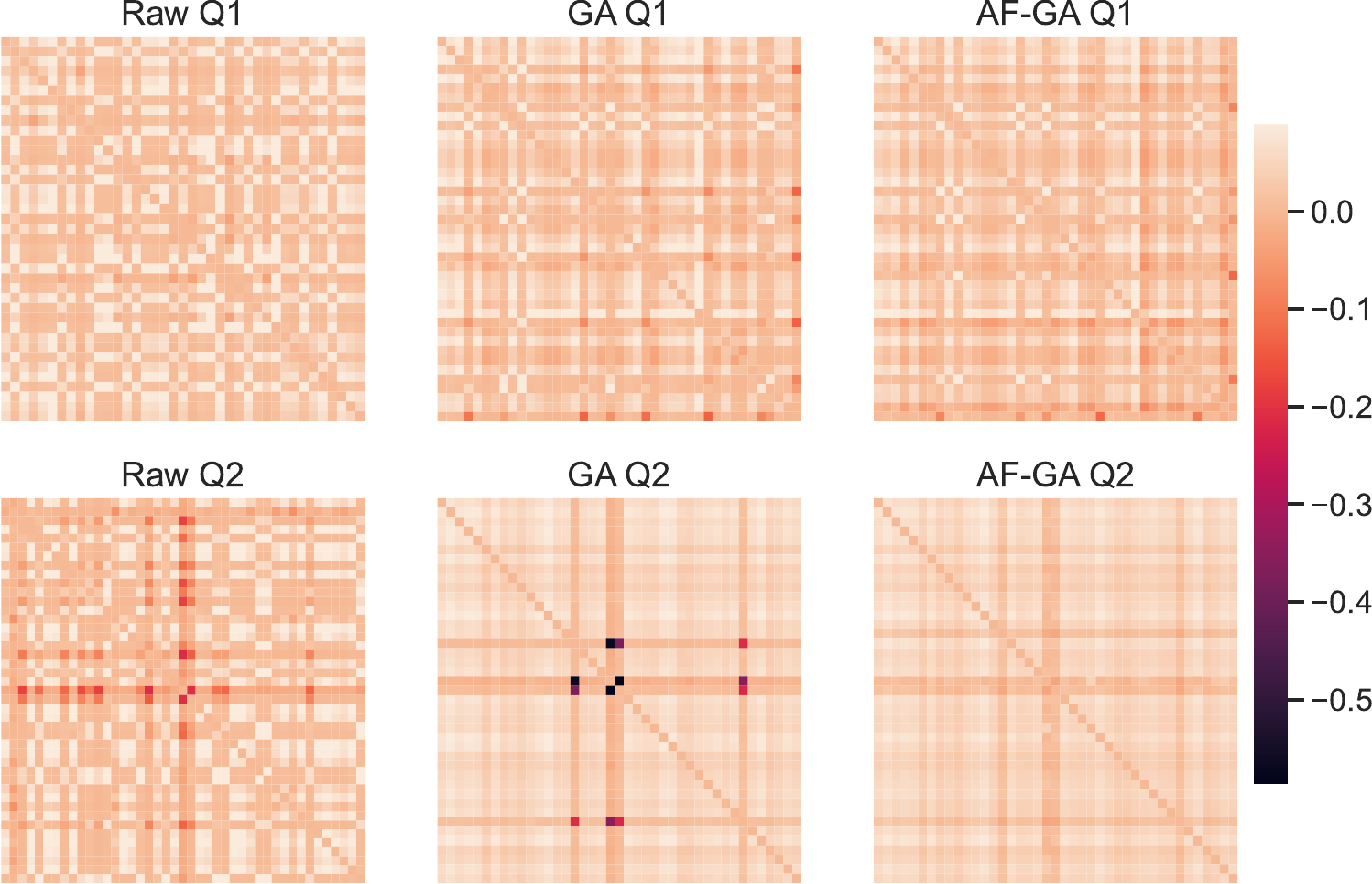}
    \caption{We plot $Q$ matrices under raw class assignment, GA assignment, and AFGA assignment. The indices of these two classrooms are 5 and 6. Two columns and rows of dark grids in Raw Q2 indicate that in classroom 6, two disruptive peers negatively influence the entire classroom to a large extent. The extremely dark grids in GA Q2 indicate that GA gives up on a few students and lets them form a clique that is almost exclusively- and negatively-influencing those in the clique. AFGA has a higher average peer effect than GA as shown by the lighter color (on average) of AF-GA plots than Raw plots. AFGA also does not suffer from the extreme inequity problem as in GA Q2. Appendix \ref{Appendix class assignment results} shows that the same patterns persist throughout various classrooms/schools.}
    \label{Realized peer effect grid plot}
\end{figure}

To assist visualizing the class assignment results for a single classroom, we define the \textbf{demeaned} peer effect $q_{ij}$ for each \textbf{pair} of student $\{i,j\}$, $q_{ij} = \Omega_{ij} \tilde{z}_j + \Omega_{ji} \tilde{z}_i$ where $\tilde{z}_i = z_i - \frac{1}{N}\sum_{j=1}^N z_j$. We collect all paired peer effect, $q_{ij}$, into a matrix $Q$, and since $q_{ij} = q_{ji}$ by definition, $Q$ is an $(N \times N)$ symmetric matrix. Technically, $\beta Q$ is each pair's predicted demeaned peer effect, but since $\beta > 0$ is a constant, we omit it. In Figure \ref{Realized peer effect grid plot}, we illustrate $Q$ as a heat map, with the term `raw' assignment referring to the initial classroom assignment policy randomly generated by the school principal.

Under the raw assignment, many students are negatively influenced by two disruptive peers as indicated by the two darker rows and columns of the Raw Q2 plot in Figure \ref{Realized peer effect grid plot}. GA separates many pairs of students who suffer from such large negative peer effects by ensuring that these two disruptive peers are popular among fewer students, as evidenced by the reduced number of dark grids in GA Q1 and GA Q2 relative to Raw Q1 and Raw Q2. Moreover, GA Q2 shows that the entire classroom consists mostly of positively influencing friend pairs. 
Unfortunately, maximizing the average peer effect inadvertently leads to severe negative peer effects for some students. A few students face extremely negative peer effects, as indicated by the very dark grids in Figure \ref{Realized peer effect grid plot} GA Q2. Since these students have to disruptive peers for $q_{ij}$ to be very negative, we can infer that GA succeeds in maximizing peer effect by predicting an exclusive friendship circle among disruptive peers. In other words, these disruptive peers are ``given up'' by GA in pursuit of a higher average peer effect.

\begin{figure}
    \centering
    \includegraphics[width = 0.5\textwidth]{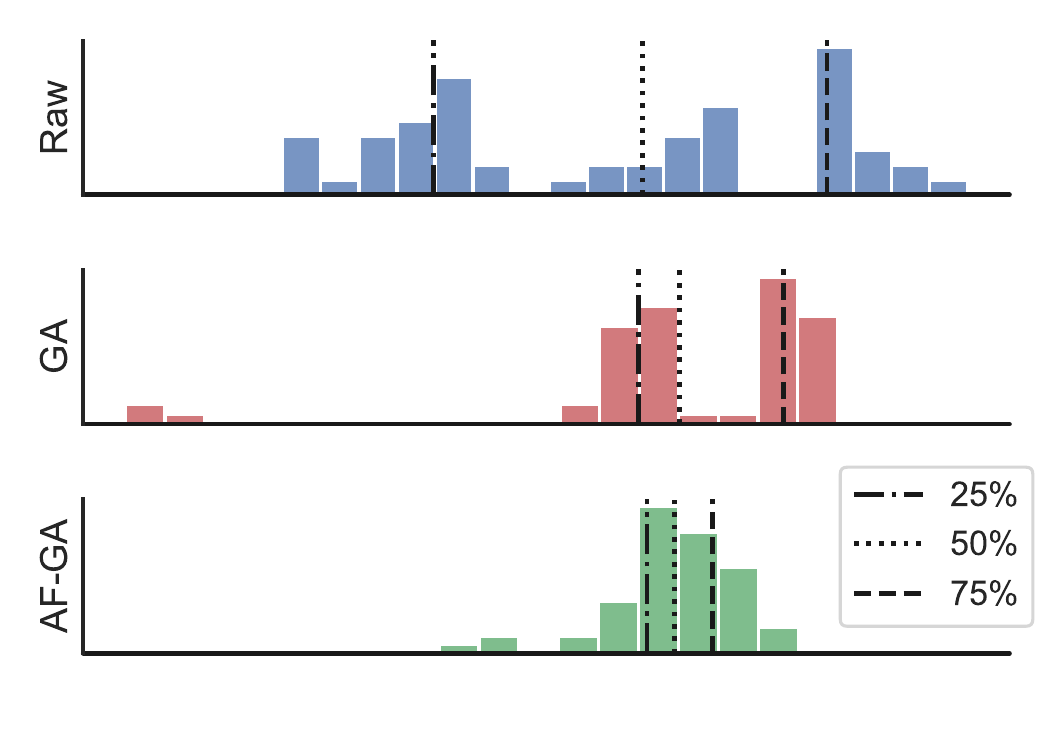}
    \caption{Histogram of $\tilde{\Omega}$, under raw, GA and AFGA class assignment policies. The dash-dotted, dotted, and dashed vertical lines mark the 25$\%$, 50$\%$, and 75$\%$ quantile of peer effects, respectively. Note that since we include classroom fixed effects in our regression model, $\tilde{\Omega}$ is translation invariant, and hence, the absolute scale of the horizontal axis is not interpretable.}
    \label{Realized_peer_effect_histogram}
\end{figure}

We deploy AFGA to address this inequity issue. AF-GA Q1 and AF-GA Q2 in Figure \ref{Realized peer effect grid plot} show a much more equitable educational outcome. None of the students suffer from extremely negative peer effects as in the GA Q2. 

In Figure \ref{Realized_peer_effect_histogram}, we compare the equity of all three policies more directly. Raw class assignment policy generates a trimodal distribution of realized peer effect. The leftmost mode is due to the strong influence of disruptive peers. GA essentially breaks the friend circle that generates the leftmost mode. However, in the process of pursuing the highest average, it ignores the peer effect for a small number of students, generating extremely inequitable educational outcomes as evidenced by the few left-tail outliers. AFGA takes into account the standard deviation of the entire distribution, outputting a much fairer policy rule.

If the school principal believes this particular policy designed by AFGA overvalues equity and sacrifices efficiency too much, he can decrease $\phi$ and $\rho$ for a more efficient outcome, however, the outcome will be less equitable. Note that the school principal can predict peer effect, $\tilde{\Omega}$, for all students before implementing the policy, hence he can always choose $\phi$ and $\rho$ based on his preference over efficiency versus equity.

\section{Conclusion} \label{Conclusion}
This study constructs a micro-founded model for predicting friendship formation and employs a novel and interpretable neural network architecture called PeerNN to estimate this model. Leveraging the predictions generated by PeerNN, we consistently estimate peer effects using a linear-in-means instrument. By combining the results of friendship formation prediction and peer effect parameter estimation, we simulate counterfactual peer effects for all students. Our work then designs an algorithmic fair genetic algorithm, outputting a class assignment policy that enhances average peer effects while maintaining fairness. 

Admittedly, implementing our framework in real life is challenging. Nevertheless, counterfactual simulation results of our framework shed light on policy recommendations. Under endogenous spillover, an efficiency-focused objective function can severely hurt the disadvantaged subpopulation. Hence, incorporating ethical considerations and equity safeguards into the policy objective should be the standard practice under such a setup.



\bibliographystyle{plainnat} 
\bibliography{name}

\appendix

\section{Friendship questionnaire} \label{Appendix Friendship questionnaire}
\begin{figure}[htb]
    \centering
    \includegraphics[width = \textwidth]{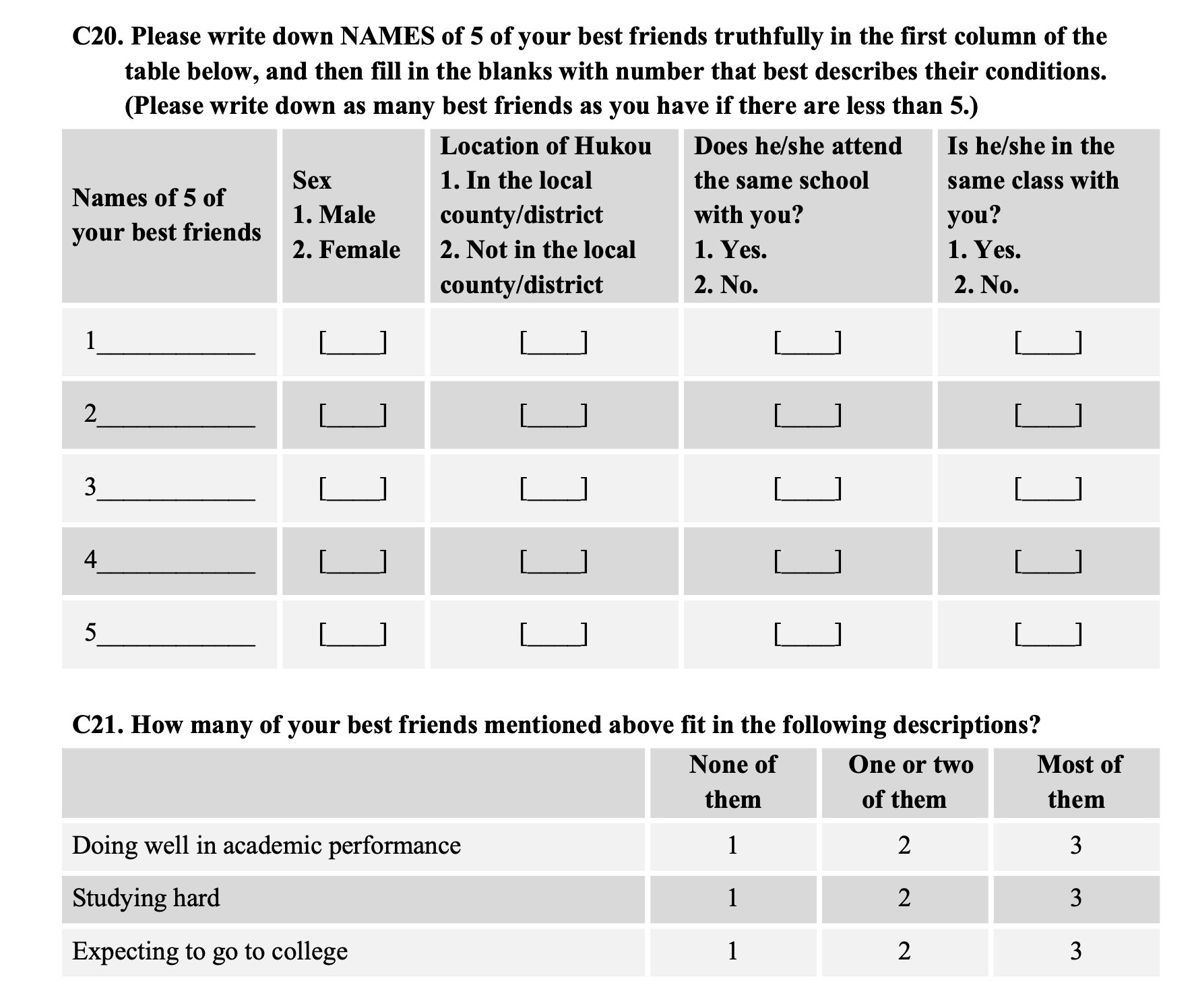}
    \caption{The friendship data is aggregated relational. The identification of network formation parameters with such data is studied by \citet{breza2020using}. We focus on friendship formation prediction instead of inference on the preference parameters. For section C21, there are nine questions in total. We do not include all the questions in section C21 in the figure. We will use students' responses to these questions ($A_f$) as the response variable for PeerNN.}
    \label{friendship survey questions}
\end{figure}

The National Survey Research Center at Renmin University of China (NSRC) provides CEPS and all information related to the dataset. The questionnaire can be found at \url{http://ceps.ruc.edu.cn/Enlish/dfiles/11184/14391213173188.pdf} (English version) and \url{http://ceps.ruc.edu.cn/__local/C/67/E7/F7BA6FBDFBC3C8B808AA0F6E0FA_905BECDA_7A84E.pdf?e=.pdf} (Chinese version). In Figure \ref{friendship survey questions}, we show a screenshot of the ARD questions from the 7th-grade survey. More ARD questions are not included in the screenshot, interested readers can use the aforementioned URLs to look at the questions. The data that we use is from the 8th grade survey, however, the ARD questions from the 8th grade survey are the same as the ARD questions from the 7th grade. We provide the screen of the 7th-grade survey as the English translation version is only available for the 7th-grade survey, not the 8th-grade survey.

\section{Examples} \label{Appendix examples section}

\begin{example} \label{example matching}
    Instead of using latent representations $\sigma(X)$, we are going to use raw data features, $X$, to illustrate the friendship demand concept. We find this simplified illustration (using $X$ instead of $\sigma(X)$) more intuitive, but it still captures the essence of the microfoundation of the model. Stage 0 (latent feature encoder) is not of the essence of the four stages in the micro-founded friendship prediction model. One can drop Stage 0 completely, it does not affect the training result much. We add Stage 0 to improve the flexibility of the neural network.

    Use two predictors: gender and whether a student did well in 6th grade to predict friendship formation. We have the data matrix (Table \ref{data matrix example}) for a classroom consisting of five students.
    
    \begin{table}
        \parbox{.48\linewidth}{
        \centering
        \begin{tabular}{c c c}
            \toprule
            & gender & did well in 6th-grade\\
            \midrule
            Adam & 1 & 1\\
            Ben  & 1 & 1\\
            Cam  & 1 & 0\\
            Debbie  & 0 & 1\\
            Emily & 0 & 0\\
            \bottomrule
        \end{tabular}
        \caption{Without encoder, $\sigma(X) = X$ contains students' predetermined characteristics: gender and whether the student did well in 6th-grade}
        \label{data matrix example}
        }
        \hfill
        \parbox{.48\linewidth}{
        \centering
        \begin{tabular}{c c c}
            \toprule
            & gender & did well in 6th-grade\\
            \midrule
            Adam & 1 & 0.5\\
            Ben  & 1 & 0.5\\
            Cam  & 0.5 & -0.5\\
            Debbie  & -1 & 0.5\\
            Emily & -0.5 & -0.5\\
            \bottomrule
        \end{tabular}
        \caption{Preference parameter $\delta$ indicates students' preference for making friends with classmates with certain characteristics}
        \label{preference parameters example}
        }
    \end{table}
    Since both input variables are binary, there are four types of students. (1) boy who did well, (2) boy who did not do well, (3) girl who did well, and (4) girl who did not do well. 
    Assume that the four types of students have the following preferences:
    \begin{itemize}
        \item Boys who did well in elementary school prefer to make friends with boys over girls and they prefer to make friends with those who did well in elementary school.
        \item Boys who did not do well in elementary school prefer to make friends with boys over girls and they prefer to make friends with those who did not do well in elementary school.
        \item Girls who did well in elementary school prefer to make friends with girls over boys and they prefer to make friends with those who did well in elementary school.
        \item Girls who did not do well in elementary school prefer to make friends with girls over boys and they prefer to make friends with those who did not do well in elementary school.
    \end{itemize}
    A possible $\delta$ matrix to describe such a preference pattern is Table \ref{preference parameters example}. Note that $\delta_i$ is a function of $X_i$ and since Adam and Ben share the same $X$ values, their preference parameters are identical.

\end{example}

\begin{example} \label{example Stage 2 and 3}
    We compute the outer product of $\delta$ and $\sigma$ from example \ref{example matching} to compute the linearized propensity score $\Upsilon_{ij}$.
    
    \begin{equation*}
        \begin{pmatrix}
            1 & 0.5 \\
            1 & 0.5 \\
            0.5 & -0.5 \\
            -1 & 0.5 \\
            -0.5 & 0.5
        \end{pmatrix}
        \begin{pmatrix}
            1 & 1 & 1 & 0 & 0 \\
            1 & 1 & 0 & 1 & 0
        \end{pmatrix}
        = \begin{pmatrix}
            1.5 & 1.5 & 1 & 0.5 & 0 \\
            1.5 & 1.5 & 1 & 0.5 & 0 \\
            0 & 0 & 0.5 & -0.5 & 0 \\
            -0.5 & -0.5 & -1 & 0.5 & 0 \\
            0 & 0 & -0.5 & 0.5 & 0 \\
        \end{pmatrix}
        = \Upsilon
    \end{equation*}

    We then set the diagonal entries of $\Upsilon$ to be negative infinity such that the probability of a student making friends with himself is 0. After that, we apply a softmax function to each row of $\Upsilon$.

    \begin{align*}
        \begin{pmatrix}
            1.5 & 1.5 & 1 & 0.5 & 0 \\
            1.5 & 1.5 & 1 & 0.5 & 0 \\
            0 & 0 & 0.5 & -0.5 & 0 \\
            -0.5 & -0.5 & -1 & 0.5 & 0 \\
            0 & 0 & -0.5 & 0.5 & 0 \\
        \end{pmatrix}
        \overset{\Upsilon_{ii} = - \infty}{\implies}
        &\begin{pmatrix}
            - \infty & 1.5 & 1 & 0.5 & 0 \\
            1.5 & - \infty & 1 & 0.5 & 0 \\
            0 & 0 & - \infty & -0.5 & 0 \\
            -0.5 & -0.5 & -1 & - \infty & 0 \\
            0 & 0 & -0.5 & 0.5 & - \infty \\
        \end{pmatrix} \\\\
        \overset{\text{row-wise softmax}}{\implies}
        &\begin{pmatrix}
            0 & 0.455 & 0.276 & 0.167 & 0.102 \\
            0.455 & 0 & 0.276 & 0.167 & 0.102 \\
            0.277 & 0.277 & 0 & 0.169 & 0.277 \\
            0.235 & 0.235 & 0.143 & 0 & 0.387 \\
            0.235 & 0.235 & 0.143 & 0.387 & 0 \\
        \end{pmatrix}
        = \Omega
    \end{align*}
\end{example}

\begin{example}\label{endogeneity example}
    Students may choose to make friends with those of similar IQ as themselves. Higher IQ also has a positive impact on students' 8th-grade cognitive ability test scores. We do not observe IQ and hence it is part of the error term. Then, if we run regression based on (\ref{second stage equation}), we may overestimate peer effect $\beta$ since we inevitably attribute some of the impact of IQ to peer effect.
    
\end{example}

\begin{example} \label{example IV propoerties}
    We continue with example \ref{endogeneity example} where IQ is an endogenous variable. 
    
    Exclusion constraint requires the instrument (average 6th-grade class rank for all classmates) to be uncorrelated with the unobserved confounder IQ. This requirement is trivially satisfied since all classroom assignments are random conditional on school choice which is a covariate included in (\ref{second stage equation}).

    Relevance constraint requires the instrument and the endogenous variable to be correlated. Both average 6th-grade class rank ($W_{cs}z_{cs}$) and friends' weighted average 6th-grade class rank ($\Omega_{cs}z_{cs}$) are dependent on all classmates' 6th-grade class rank ($z_{cs}$). Therefore, the relevance constraint is satisfied.
\end{example}


\section{Explanation for the loss function} \label{Appendix loss function}

\subsection{Fitted value's MSE} \label{Appendix MSE}
CEPS provides aggregated relational data. We design Stage 4 of PeerNN in accordance with this data feature. If researchers have network linkage data, this stage can be simplified.

Given a $(N \times N)$ matrix $\Omega$ whose rows are multinomial distributions, $\Omega_{ij}$ is the probability of students $i$ considering student $j$ as a friend. The diagonal of $\Omega$ is 0. Each row of $\Omega$ sums to 1. If we knew the linkage data, we could do a classical maximum likelihood estimation. In the case of aggregate relational data, we can randomly draw friends for all students based on $\Omega$ \textbf{without replacement}. 
However, this gradient-estimation-based method can be computationally intensive as $\Omega$ is a large probability matrix. We make two compromises to reduce the computation complexity. Nevertheless, the out-of-sample performance of PeerNN beats the linear-in-means model by a long shot, hence, we feel comfortable with making those compromises. Future research may build more reasonable micro-founded models to predict friendship formation. 

\subsubsection{Compromise 1} 
We let students draw best friends \textbf{with replacement}. In reality, when students answer questions in Figure \ref{friendship survey questions}, they are unlikely to put down the same friend twice. However, a random draw with replacement does not rule out such a possibility. Nevertheless, this compromise exponentially decreases the computational burden. We show that with the following scenario.

Assume there is a class consisting of $N$ students, and all students report \textit{five} best friends. Let $V_i$ be the randomly drawn best friends vector for student $i$ \textbf{without replacement}. $V_i$ is of length $N$, the ith entry of $V_i$, $V_{ii} = 0$ because $\Omega_{ii} = 0$, and \textit{five} of the entries of $V_i$ is 1 while the rest of the entries are 0. We break MSE into bias square and variance. In order to compute the bias square, we need to compute the expectation of $V_i$. For the first friend, computing expectation is straightforward $E[V_{i(1)}] = \Omega_i$. Computing for the expectation of the first two friends is exponentially more computationally burdensome.
\begin{align*}
    E[V_{i(1)} + V_{i(2)}] &= E[V_{i(1)} + E[V_{i(2)} | V_{i(1)}]] \\
    &= \Omega_i + E[\frac{\Omega_i - V_{i(1)} \circ \Omega_i}{1- V_{i(1)}\cdot \Omega_i}]\\
    &= \Omega_i + \sum_{j \neq i}^{N} \Omega_{ij} \frac{\Omega_i - e_j \circ \Omega_i}{1- e_j\cdot \Omega_i}
\end{align*}
$e_j$ is a vector of zeros except for the jth entry being 1. $\circ$ and $\cdot$ denote elementwise product and dot product, respectively. Using the same method as in the chain of equality (law of iterated expectation), we can extend the result to the first three draws of best friends without replacement.
\begin{equation*}
    E[V_{i(1)} + V_{i(2)} + V_{i(3)}] = \Omega_i + \sum_{j \neq i}^{N} \Omega_{ij} \frac{\Omega_i - e_j \circ \Omega_i}{1- e_j\cdot \Omega_i} + \sum_{k \notin {i,j}}^{N}\sum_{j \neq i}^{N} \Omega_{ik}\Omega_{ij} \frac{\Omega_i - e_j \circ \Omega_i - e_k \circ \Omega_i}{1 - e_j\cdot \Omega_i - e_k\cdot \Omega_i}
\end{equation*}
It is clear that as the number of drawn best friends that we consider increases, the computational complexity increases exponentially. For the case that we consider where all students report 5 friends, the computational complexity for computing the expectation of $V_i$ only is $O(N^5)$. For each student, the computational complexity is $O(N^4)$ and there are $N$ students that we need to compute bias for. Moreover, since CEPS provides aggregate relational data only, our loss function will be even more complex than computing the expectation of $V_i$ only. As such, we make the `with replacement' compromise. We will show that with compromises 1 and 2, the complexity of the MSE component of the loss function is reduced to $O(N^3)$.



\subsubsection{Compromise 2} 
We use five linear functions to \textbf{approximate the correspondence} between drawn friends' characteristics and students' evaluation of their friends. The survey design of CEPS forces surveyed students to report their responses in $\{1,2,3\}$ as depicted in Figure \ref{friendship survey questions}. The prediction that we make should match students' responses about their best friends. 

Given $\Omega$ and students' self-evaluation $A_s$, we can randomly draw friends for students and predict, for example, how many of their friends study hard based on randomly drawn friendship vector $V_i$ and the column in $A_s$ which corresponds to an indicator of whether a student studies hard or not. The product of $V_i A_s$ informs us how many of the best friends whom student $i$ reports exhibit a trait such as studying hard. Note that $V_i A_{sq} \in \{0,1\dots, B_i\}$ where $B_i$ is the number of reported best friends up to 5. Hence, there needs to be a mapping from $(V_i A_{s}, B_i)$ to students' evaluation of their best friends, $A_{fi}$, we dropped the subscript for question $q$ since the mapping is the same for all questions. $A_s$ and $A_{fi}$ are both $(N \times 10)$ matrices. Table \ref{table arbitrary g} documents the \textbf{correspondence or multi-valued function}, $\mathcal{G}: \mathcal{B} \times \mathcal{V} \times \mathcal{A} \to \{1,2,3\}$.

\begin{table}[htb]
    \centering
    \begin{tabular}{c c c c c c c}
        & $V_{i} \cdot A_{sq} = 0$ & $V_{i} \cdot A_{sq} = 1$ & $V_{i} \cdot A_{sq} = 2$ & $V_{i} \cdot A_{sq} = 3$ & $V_{i} \cdot A_{sq} = 4$ & $V_{i} \cdot A_{sq} =5$  \\
      $ B_i = 1$ & $\{1\}$ & $\{2,3\}$ & $\times$ &$\times$& $\times$ & $\times$\\
      $ B_i = 2$ & $\{1\}$ & $\{2\}$ & $\{2,3\}$ &$\times$&$\times$ & $\times$\\
      $ B_i = 3$ & $\{1\}$ & $\{2\}$ & $\{2,3\}$ & $\{3\}$ &$\times$&$\times$\\
      $ B_i = 4$ & $\{1\}$ & $\{2\}$ & $\{2\}$ & $\{3\}$ & $\{3\}$ & $\times$\\
      $ B_i = 5$ & $\{1\}$ & $\{2\}$ & $\{2\}$& $\{3\}$ & $\{3\}$ &$\{3\}$\\
    \end{tabular}
    \caption{Students' responses to friendship questionnaire may cause ambiguity. Given $B_i$ and $V_{i} \cdot A_{sq}$, we cannot uniquely pin down students' responses to the friendship questionnaire.}
    \label{table arbitrary g}
\end{table}

\begin{figure}[htb]
    \centering
    \includegraphics[width = 0.6\textwidth]{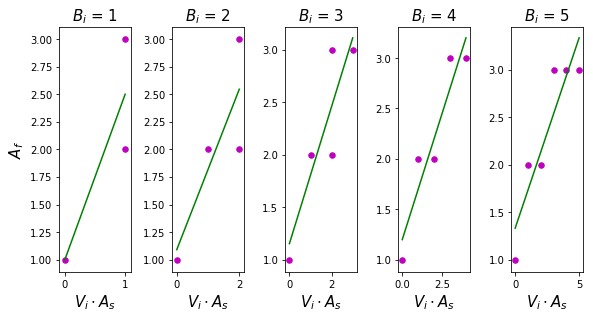}
    \caption{Five linear functions $g$ approximate the correspondence $\mathcal{G}$. The dots are the output values of $\mathcal{G}$ (i.e. values in Table \ref{table function g}). The lines are the least square linear best fit of those dots. We need $g$ to be linear in $V_i$ and hence, allow the intercept of slope to vary with $B_i$.}
    \label{figure function g}
\end{figure}

There are two problems with $\mathcal{G}$. One, its output can be multi-valued and we cannot uniquely pin down students' evaluation of their friends even if we know who their friends are. Two, it is nonlinear and renders ${\rm Bias}^2$ and ${\rm Var}$ to have no closed forms. We approximate $\mathcal{G}$ with five linear functions. The five linear functions, named $g$, are depicted in Figure \ref{figure function g}. $g_{B_i}(V_i,A_s) = a_{B_i} V_i\cdot A_s + b_{B_i}$ is linear in $V_i$, the only random variable in function $g$. We collect the intercept and slope of these five linear functions in Table \ref{table function g}.

\begin{table}[htb]
    \centering
    \begin{tabular}{c c c c c c}
    \toprule
         $B_i$ & 1 & 2 & 3 & 4 & 5 \\
         \midrule
         $b_{B_i}$&1. & 1.090 &1.154 &1.2 & 1.333\\
         $a_{B_i}$&1.5 &  0.727 & 0.654 & 0.5  &0.4\\
         \bottomrule
    \end{tabular}
    \caption{Intercept and slope coefficients $g$.}
    \label{table function g}
\end{table}


\subsubsection{Decomposition of MSE}
We can work out the decomposition as a bias-variance formula. 
\begin{align*}
    & E[\overbrace{(g(V_i A_{s}, B_i)}^{Q \times 1 \text{vector}}
     - A_{fi})^{\top}(g(V_i A_s, B_i) - \overbrace{A_{fi}}^{Q \times 1 \text{vector}})] \\
    =& E[(g(V_i A_s, B_i) - E[g(V_i A_s, B_i)] + E[g(V_i A_s, B_i)] - A_{fi})^{\top}\\
    &(g(V_i A_s, B_i) - E[g(V_i A_s, B_i)] + E[g(V_i A_s, B_i)] - A_{fi})] \\
    =& \underbrace{(E[g(V_i A_s, B_i)]- A_{fi})^{\top}(E[g(V_i A_s, B_i)]- A_{fi})}_{\text{bias square}} + \\
    &\underbrace{E[(g(V_i A_s, B_i) - E[g(V_i A_s, B_i)])^{\top}(g(V_i A_s, B_i) - E[g(V_i A_s, B_i)])]}_{\text{`variance'}} +\\
    & E[2\underbrace{(E[g(V_i A_s, B_i)] - A_{fi})^{\top}}_{\text{constant}}\underbrace{( g(V_i A_s, B_i) - E[g(V_i A_s, B_i)])}_{\text{zero expectation}}]
\end{align*}

Next, we show how to relate the `variance' term to the variance of $g(V_i A_s,B_i)$.
\begin{align*}
    &E[(g(V_i A_s, B_i) - E[g(V_i A_s, B_i)])^{\top}(g(V_i A_s, B_i) - E[g(V_i A_s, B_i)])] \\
    =&E[tr((g(V_i A_s, B_i) - E[g(V_i A_s, B_i)])^{\top}(g(V_i A_s, B_i) - E[g(V_i A_s, B_i)]))]\\
    =&E[tr((g(V_i A_s, B_i) - E[g(V_i A_s, B_i)])(g(V_i A_s, B_i) - E[g(V_i A_s, B_i)])^{\top})]\\
    =&tr(E[(g(V_i A_s, B_i) - E[g(V_i A_s, B_i)])(g(V_i A_s, B_i) - E[g(V_i A_s, B_i)])^{\top}])\\
    =&tr({\rm Var}(g(V_iA_s,B_i)))
\end{align*}

\subsubsection{Closed-form expressions for \texorpdfstring{${\rm Bias}^2$}{O} and \texorpdfstring{${\rm Var}$}{0}}
With compromise 1 and 2, we can simplify $E[g(V_i A_s, B_i)]$ and ${\rm Var}(g(V_i A_s, B_i))$ into the following closed-form expressions:
\begin{align*}
    E[g(V_i A_s, B_i)] &= E[a_{B_i}V_iA_s + b_{B_i}] = a_{B_i}\Omega_iA_s + b_{B_i} \\
    {\rm Var}(g(V_i A_s, B_i)) &=  {\rm Var}(a_{B_i}V_iA_s) = a^2_{B_i}A_s^{\top}{\rm Var}_{\Omega_i}A_s \\
    \text{where} \quad 
    {\rm Var}_{\Omega_i} = {\rm Var}(V_i) &= \begin{cases}
        {\rm Var}(V_{ijj}) = B_i \Omega_{ij} (1 - \Omega_{ij}) \quad \text{for $j \in \{1,2,\dots,N\}$.}\\
        {\rm Var}(V_{ijk}) = - B_i \Omega_{ij} \Omega_{ik} \quad \text{when $j \neq k$.}
    \end{cases}
\end{align*}
We estimate the ${\rm Bias}^2$ and ${\rm Var}$ in the loss function for one classroom as follows:
\begin{align*}
    {{\rm Bias}}^2 =& \frac{1}{NQ}\sum_{i=1}^{N} (a_{B_i}\Omega_i A_s + b_{B_i} - A_{fi})^{\top}(a_{B_i}\Omega_i A_s + b_{B_i} - A_{fi})  \\
    {{\rm Var}} =& \frac{1}{NQ}tr(\sum_{i=1}^{N} a_{B_i}^2 A_s^{\top} {\rm Var}_{\Omega_i} A_s)
\end{align*}
where $N$ is the number of students in the classroom and $Q$ is the number of questions that are asked in the survey ($Q=10$). 
For one student, the computational complexity for MSE is the same as that for ${\rm Var}_{\Omega_i}$ which is $O(N^2)$. We do it for $N$ students, thus, the total complexity is $O(N^3)$.

\subsection{Downweighing variance} \label{Appendix downweighing variance}
The benefit of downweighing variance can be intuitively explained with social network density. The classroom network exhibits a dense nature: students report more than 1 friend (most of the students report 5 friends which is the maximum number of friends that they can report). When $(\mu,\kappa,\lambda) = (1,0,0)$, we observe that fitted $\Omega$ matrix predicts that all friendship forming probability is concentrated on one boy and one girl, all boys choose one boy as their only friend; all girls choose one girl as their only friend. This happens because extremely concentrated friendship forming probability $\Omega$ results in close to zero variance, indicating that variance is overweighted. Consequently, the neural network predicts an almost deterministic friendship formation rule, with each row having one entry being close to one and other entries being close to zero. This violates the dense nature of classroom social networks, therefore, we downweigh variance to account for social network density.

\subsection{Using fitted \texorpdfstring{$\Omega$}{O} to select \texorpdfstring{$(\mu,\kappa,\lambda)$}{ukl}} \label{Appendix tuning parameter selection}
By leveraging common knowledge of social network structure, we can avoid cross-validation when tuning hyperparameters $(\mu,\kappa,\lambda)$. We examine whether the $\Omega$ matrix generated by a combination of $(\mu,\kappa,\lambda)$ satisfies the commonly known properties of social networks. Such examination uses information from train data exclusively and does not require any validation data.

For example, as mentioned in Appendix \ref{Appendix downweighing variance}, $(\mu,\kappa,\lambda) = (1,0,0)$ is unreasonable as the $\Omega$ matrix generated by this combination of tuning parameters violates the dense nature of classroom network. 

Another example is when $(\mu,\kappa,\lambda) = (1, 60, 100)$, $\Omega$ is almost a constant within each column: all boys are equally influenced by their boy classmates; all girls are equally influenced by their girl classmates. This violates the clustering property of social networks: students should exhibit varying degrees of influence on different groups of students.

We pick $(\mu,\kappa,\lambda) = (0.2,3,3)$. The choice of hyperparameters might seem ad hoc; however, one can formalize measures of gender homophily, node centrality, and clustering to optimize these hyperparameter choices more systematically. Such a tuning method is time-efficient because we only need to fit the model once for a particular choice of $(\mu,\kappa,\lambda)$ in comparison to $k$-fold cross-validation which requires $k$ instances of the fitting.



\section{Linear-in-means model} \label{Appendix linear in means}
The conventional econometric model for peer effect measurement is the linear-in-means model

 \begin{equation*}
        y_{ics} = \beta \underbrace{\frac{\sum_{j=1,j\neq i}^{N_{cs}} z_{jcs}}{N_{cs}-1}}_{\text{Exclusive means}} + X_{ics}\gamma + \underbrace{\theta_s}_{FE} + \underbrace{\mu_{cs}}_{RE}  + \epsilon_{ics}
    \end{equation*}
where $y_{ics}$ is student $i$'s 8th-grade cognitive ability in class $c$ school $s$; $N_{cs}$ is the class size of classroom $c$ from school $s$; $z_{jcs}$ is classmate's 6th-grade class rank in class $c$ school $s$; $X_{ics}$ is additional controls for student $i$ such as gender, age, ethnic group, and parents' education; $\theta_s$ is school fixed effects and $\mu_{cs}$ is class random effects. We can rewrite linear-in-means model as the $y_{cs} = \beta W_{cs} z_{cs} + X_{cs} \gamma + \theta_s +\mu_{cs} + \epsilon_{ics}$ where

\begin{equation*}
    W_{cs} = \begin{pmatrix}
        0 &\frac{1}{N_{cs}-1} &\frac{1}{N_{cs}-1} &\dots & \dots &\frac{1}{N_{cs}-1} \\
        \frac{1}{N_{cs}-1} &0 &\frac{1}{N_{cs}-1} &\dots &\dots &\frac{1}{N_{cs}-1} \\
        \vdots & \vdots &\vdots & \vdots &\vdots &\vdots\\
        \frac{1}{N_{cs}-1} &\frac{1}{N_{cs}-1} &\frac{1}{N_{cs}-1} &\dots & \dots & 0
    \end{pmatrix}.
\end{equation*}
Linear-in-means model implicitly assumes that all students exert the same amount of peer influence on all their classmates as evidenced by the $W_{cs}$ matrix. The structural equation that contains the parameter of interest $\beta$ (i.e. uation (\ref{second stage equation})) replaces $W_{cs}$ with $\Omega_{cs}$. The microfoundation for specification of uation (\ref{second stage equation}) is the following.

\subsection{Microfoundation of (\ref{second stage equation})}

In our model, we assume that students continuously update their best friend, and momentarily, a student receives peer effect exclusively from his best friend. Therefore, the peer effect a student receives from one of his classmates is proportional to the probability of him considering that classmate to be his best friend. 

As illustrated by example \ref{endogeneity example}, ${\rm Cov}(\Omega_{cs}, \epsilon_{cs}) \neq 0$. In economics, this problem is called endogeneity. We address endogeneity by constructing an instrumental variable which happens to be $W_{cs}z_{cs}$. 

\subsection{Comparison between linear-in-means and friendship-weighted specifications}
The two $\beta$s in (\ref{linear in means specification}) and (\ref{second stage equation}) have different interpretations. For linear-in-means model, $\beta$ in (\ref{linear in means specification}) is the marginal effect of middle school \textbf{classmates' average} 6th-grade class quantile on the outcome; for friendship-weighted specification, $\beta$ in (\ref{second stage equation}) is the marginal effect of middle school \textbf{friends' weighted average} 6th-grade class quantile on the outcome. The former has very little policy implication on the average peer effect since the linear-in-means model essentially restricts how much total peer effect each student can exert on other classmates. Our friendship-weighted specification relaxes this restriction: a popular student is allowed to exert more influence on all his classmates than another less popular student. Students may also have different levels of influence on different subgroups, for example, in our empirical results, boys influence boys more and girls influence girls more. This allows us to devise average-maximizing class assignments.

\section{Genetic algorithm used to optimize class assignment policy} \label{Appendix GA algorithm}
We constrain that for a school which has fewer girls than boys, it has to assign at least 35\% and at most 65\% of the girls to one classroom; for a school which has fewer boys than girls, it has to assign at least 35\% and at most 65\% of the boys to one classroom. 

Some of the parameters that we use for our GA design include (1) the number of iterations is 150, (2) the mutation probability is 5\%, and (3) the number of swaps for each iteration is 100.

Let $C$ denote the set of all students that the principal is going to divide into classrooms. Let $C_1$ and $C_2$ denote the set of students from the first and second classrooms, respectively. Note that $(C_1, C_2)$ fully pins down the class assignment policy. We constrain that $-1 \leq \lvert C_1 \rvert - \lvert C_2 \rvert \leq 1$, ensuring that the classrooms are of comparable sizes. For simplicity, we first illustrate two functions that we repeatedly use in our GA design with Algorithm \ref{check function} and Algorithm \ref{swap function}.

          

\begin{algorithm}[htb]
\caption{check function}
\label{check function}

     \textbf{Input}: $C_1, C_2, N_b, N_g, j \in\{b,g\}$
    
     \textbf{Purpose}: Validate whether class assignment meets gender ratio constraints
     
    Function: \textbf{check}($C_1, C_2$)
        
        \If{$0.35 \cdot N_{j} \leq \sum_{s \in (C_1)} \mathbbm{1}_{\{ s \text{ is gender $j$}\}} \leq 0.65 \cdot N_{j}$}
             {\textbf{return} 1}
        \Else
             {\textbf{return} 0}

\end{algorithm}


\begin{algorithm}[htb]
\caption{swap function}
\label{swap function}

     \textbf{Input}: $C_1, C_2, CP$
    
     \textbf{Purpose}: Swap students between two classrooms
     
    Function: {swap}({$C_1, C_2, CP$})
        
        \If{$CP \neq ({\rm NULL},{\rm NULL})$}
             {$CP_1, CP_2 \gets CP$ 
             
             $C_1 \gets (C_1 \setminus \{CP_1\}) \cup \{CP_2\}$
             
             $C_2 \gets (C_2 \setminus \{CP_2\}) \cup \{CP_1\}$}
        
        \textbf{return} $C_1, C_2$

\end{algorithm}

\begin{algorithm} \small
    \caption{GA used to optimize class assignment policy initialization}\label{GA algorithm}
        \KwIn{$C$, \hfill Set $C$ consists of all students the principal needs to divide into two classrooms}
         $N_b \gets \sum_{s \in C} \mathbbm{1}_{\{ s \text{ is boy}\}}$
         
         $N_g \gets \sum_{s \in C} \mathbbm{1}_{\{ s \text{ is girl}\}}$
         
         $N \gets N_b + N_g$ \hfill {Alternatively, $N = \lvert C \rvert$}
         
         $j = \argmin_j \{N_j\}_{j\in\{b,g\}}$  
         
        \Repeat{check$(C_1,C_2,N_b,N_g,j) = 1$ \hfill {See Algorithm \ref{check function} in Appendix \ref{Appendix GA algorithm}}}{
            $v \gets$ Randomly draw $\lfloor {\frac{\lvert C \rvert}{2}} \rfloor$ students from $C$ without replacement and uniform probability
            
             $C_1 \gets C[v]$ 
             
             $C_1 \gets C \backslash C_1$ {This initialization ensures that $-1 \leq \lvert C_1 \rvert - \lvert C_2 \rvert \leq 1$}
        }
             
         L = 150
         
         M = 100
        
         $PH = {\rm matrix}(0,L,N)$ \hfill{Record all the chosen class assignment policies}
         
         $FSH = {\rm rep}(0,L)$ \hfill{Record all fitness scores of the chosen policies}

    


        \For{$l$ in $1:L$}

                

              $b \gets B \sim {\rm Bernoulli}(0.05)$
              
            \Repeat{check$(C_1',C_2',N_b,N_g,j) = 1$}{
            \If{b = 1} {
                $t_{1} \gets T_{1} \sim {\rm Multinomial}(n = 1,{\rm prob} = {\rm rep}(\frac{1}{\lvert C_1 \rvert}, C_1))$
                
                     $t_{B} \gets T_{B} \sim {\rm Multinomial}(n = 1,{\rm prob} = {\rm rep}(\frac{1}{\lvert C_2 \rvert}, C_2))$
                     
                     $CP \gets (C_1[t_1], C_2[t_2])$ 
                    %
                     $C_1',C_2' \gets {\rm swap}(C_1,C_2,CP)$ \hfill {See Algorithm \ref{swap function} in Appendix \ref{Appendix GA algorithm}}
            }
            }

            \Else {$FS = {\rm rep}(0,M)$
                
                 $v_1 \gets V_1 \sim {\rm Multinomial}(n = M,{\rm prob} = {\rm rep}(\frac{1}{\lvert C_1 \rvert}, C_1))$ 
    
                 $v_2 \gets V_2 \sim {\rm Multinomial}(n = M,{\rm prob} = {\rm rep}(\frac{1}{\lvert C_2 \rvert}, C_2))$ 
                
                 $CS \gets (C_1[v_1],C_2[v_2])$
                {$CS$ denotes M candidate swaps}}

                \For{$m \in 1:M$}{ 
                     $(C_1',C_2')= {\rm swap}(C_1,C_2,CS_m)$  \hfill {$CS_m$ is the $m$th candidate swap}
                     
                    \If {check$(C_1',C_2',N_b,N_g,j) = 1$} {$FS[m] \gets  {\rm fit}(C_1', C_2')$ \hfill{`fit' is either (\ref{Fit_GA}) or (\ref{Fit_AFGA})}}
                    \Else {$FS[m] \gets -\infty$}}

                 $bp \gets \argmax_{m} FS[m]$
                 
                \If{$FS[bp] > FSH[l-1]$} {$CP \gets CS_{bp}$}
                     
                \Else{$CP \gets ({\rm NULL},{\rm NULL})$} 

             $(C_1,C_2) \gets {\rm swap}(C_1,C_2,CP)$
             
             $PH[l,] = (C_1,C_2)$
             
             $FSH[l] = {\rm fit}(C_1,C_2)$

         $bp \gets \argmax_l FSH[l]$
         
         $C_1,C_2 \gets PH[bp,]$
    
\end{algorithm}

    
        
        
        

\section{Friendship formation pattern} \label{Appendix friendship formation pattern}
We show that the friendship formation patterns that we describe in our main text are the same throughout different classroom networks with Figure \ref{heat maps FIM 8 and 2}. First, the gender homophily effect is the dominant predictor for friendship formation. Boys almost exclusively make friends with boys and girls make friends with girls. Second, the centrality pattern is clear in the heat maps. Some individual students are extremely influential to other students. Third, students have varying degrees of influence on other students in the same classroom. Even within the same gender, one student is considered by other students as best friends with different probabilities.
\begin{figure}[htb]
\centering
\subfloat[Friendship heat map for classroom 152]{\label{fig:a}\includegraphics[width=.45\linewidth]{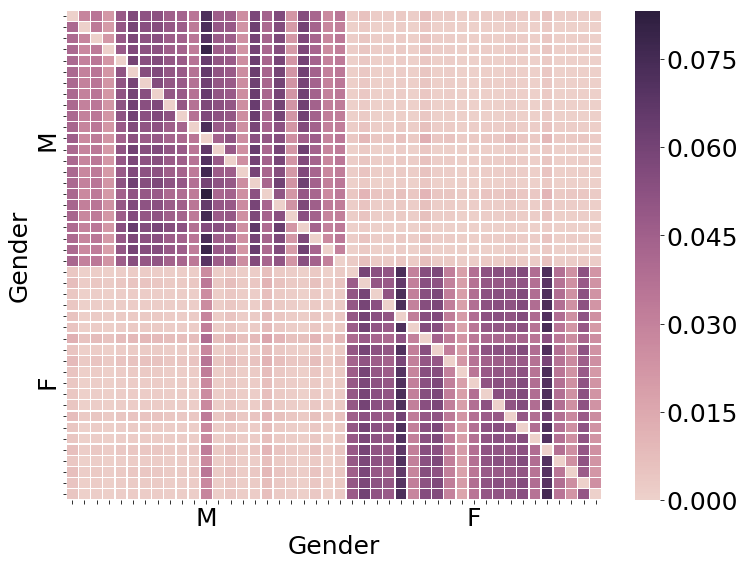}}
\subfloat[Friendship heat map for classroom 351]{\label{fig:b}\includegraphics[width=.45\linewidth]{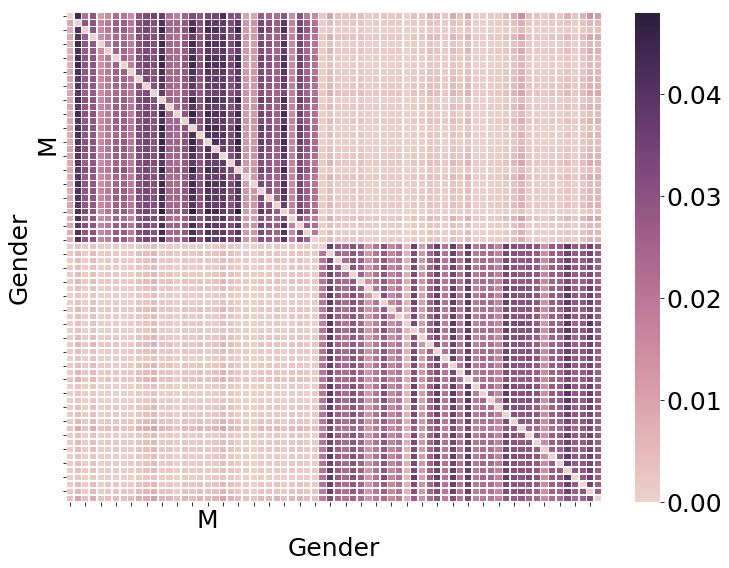}}

\caption{Heat maps for two classrooms to demonstrate that PeerNN prediction aligns with known property of social network: (1) gender homophily (2) presence of central nodes (3) heterogeneous popularity across different students}
\label{heat maps FIM 69 and 138}
\end{figure}

In Algorithm \ref{prediction error algorithm}, we describe how we compute the prediction error for the PeerNN model with test data. In the algorithm, we denote function $G$ as a modification of correspondence $\mathcal{G}$, where if $\mathcal{G}$ outputs multiple outcomes, function $G$ takes the average of them; otherwise, $\mathcal{G}$ and $G$ share the same output. For example, when $V_i \cdot A_s = 1$ and $B_i = 1$, $\mathcal{G}$ outputs $\{2,3\}$ as shown by Table \ref{table arbitrary g}, then $G$ outputs 1.5. 

\begin{algorithm}
    \caption{Computing prediction error for question $q$}\label{prediction error algorithm}

         \textbf{Input}: $\Omega,B_i,A_{sq},A_{fq}$
         
         \textbf{Output}: Prediction error
         
         ${\rm PE}_q = 0$
         
         $R = 1000$
         
        \For{$i$ in $1:N$}{
             $\Xi_i \gets \Omega_i$
             
              $V \gets {\rm rep}(0,N)$ 
              
              $d \gets {\rm rep}(0,R)$
            \For{$j$ in $1:R$}{
                \For{$j$ in $1:B_i$}
                     {$v \gets \nu \sim {\rm Multinomial}(n = 1, {\rm prob} = \Xi_i)$
                     
                     $\Xi_i \gets \frac{\Xi_i - v \circ \Xi_i}{1 - v \cdot \Xi_i}$
                     
                     $V \gets V + v$}
            
                 $d[j] \gets G(V\cdot A_{sq},B_i) - A_{fq}$}
    
            $ {\rm PE}_q \gets {\rm PE}_q + d \cdot d$}

\end{algorithm}
Algorithm \ref{prediction error algorithm} describes how prediction error is computed for a classroom with $N$ students. We sum prediction error over all classrooms in the test set.


\section{Baltagi IV, BVK IV, and full heterogeneous analysis results}
See Table \ref{Baltagi and BVK} and Table \ref{Hetero effects full}.

\setlength{\tabcolsep}{10pt}
\begin{table}[htbp]\centering
\def\sym#1{\ifmmode^{#1}\else\(^{#1}\)\fi}
\begin{tabular}{lcc}
\toprule
& Baltagi IV & BVK IV \\
\midrule
Peer's Rank & 0.975\sym{***} &  1.075\sym{***} \\
 & (0.268) & (0.270) \\
Own Rank & 0.997\sym{***} & 0.995\sym{***} \\
& (0.038) & (0.038) \\
Age & -0.012\sym{***} &  -0.012\sym{***} \\
& (0.001) & (0.001) \\
Sex & -0.011 & -0.011  \\
& (0.017) & (0.017) \\
Father's education & 0.020\sym{***} & 0.020\sym{***} \\
& (0.006) & (0.006) \\
Mother's education & 0.005 &  0.005 \\
& (0.006) & (0.006) \\
Ethnic nationality & 0.012 &  0.012 \\
& (0.040) & (0.040) \\
\midrule
Number of observations  &  5,860 &  5,860 \\
Adjusted \(R^{2}\) & 0.380 & 0.379 \\
\midrule
Results from first stage\\
Peer's Rank& 0.905\sym{***}   & 0.905\sym{***}      \\
& (0.015)  &  (0.015)    \\
Adjusted \(R^{2}\)& 0.755 & 0.755 \\
\midrule
School FE & YES & YES \\
Class RE & YES & YES \\
\bottomrule
\multicolumn{2}{l}{\footnotesize Standard errors in parentheses}\\
\multicolumn{2}{l}{\footnotesize \sym{*} \(p<0.10\), \sym{**} \(p<0.05\), \sym{***} \(p<0.01\)}\\
\end{tabular}\\
\caption{The Peer Effects of Sixth Grade Rank on Cognitive Test Score}
\label{Baltagi and BVK}
\end{table}

\section{Class assignment results}\label{Appendix class assignment results}
\begin{figure}[htb]
     \centering
     \includegraphics[width = 0.85\textwidth]{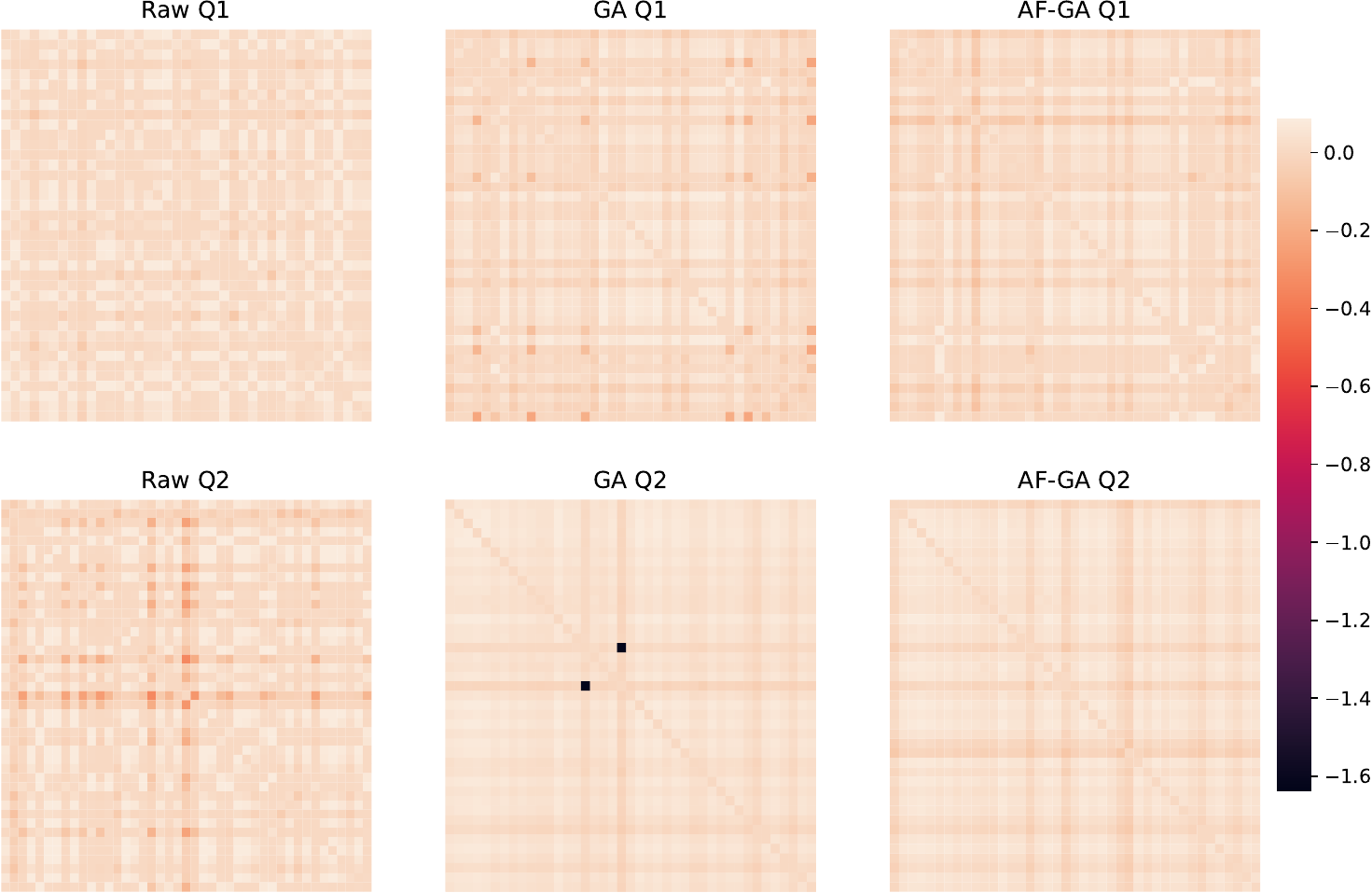}
     \caption{Classroom 5 and 6, $\phi = \rho = 1$}
     \label{grid_56_1}
 \end{figure}

\begin{figure}[htb]
     \centering
     \includegraphics[width = 0.85\textwidth]{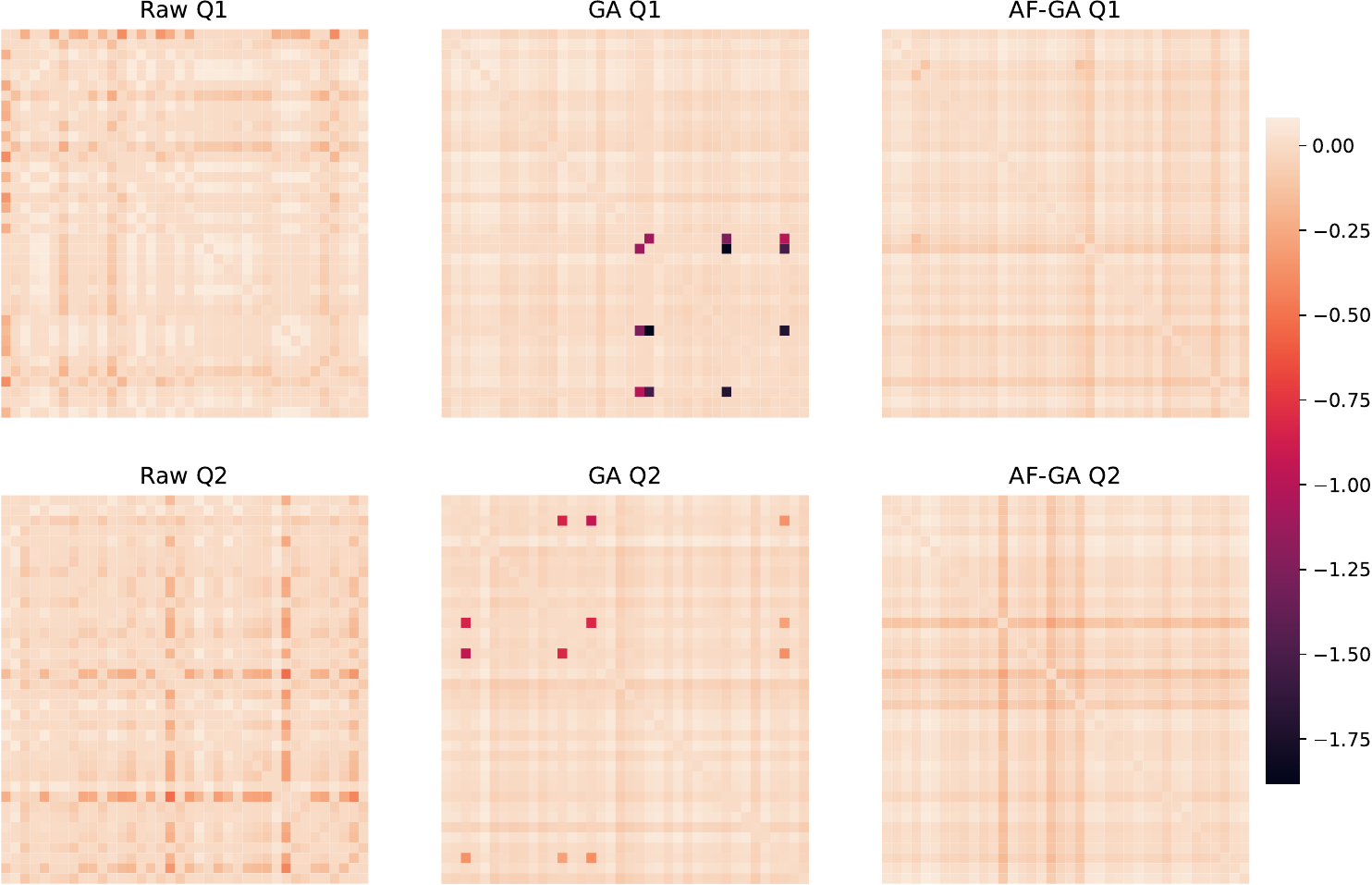}
     \caption{Classroom 25 and 26, $\phi = \rho = 2$}
     \label{grid_2526_2}
 \end{figure}

\begin{figure}[htb]
     \centering
     \includegraphics[width = 0.85\textwidth]{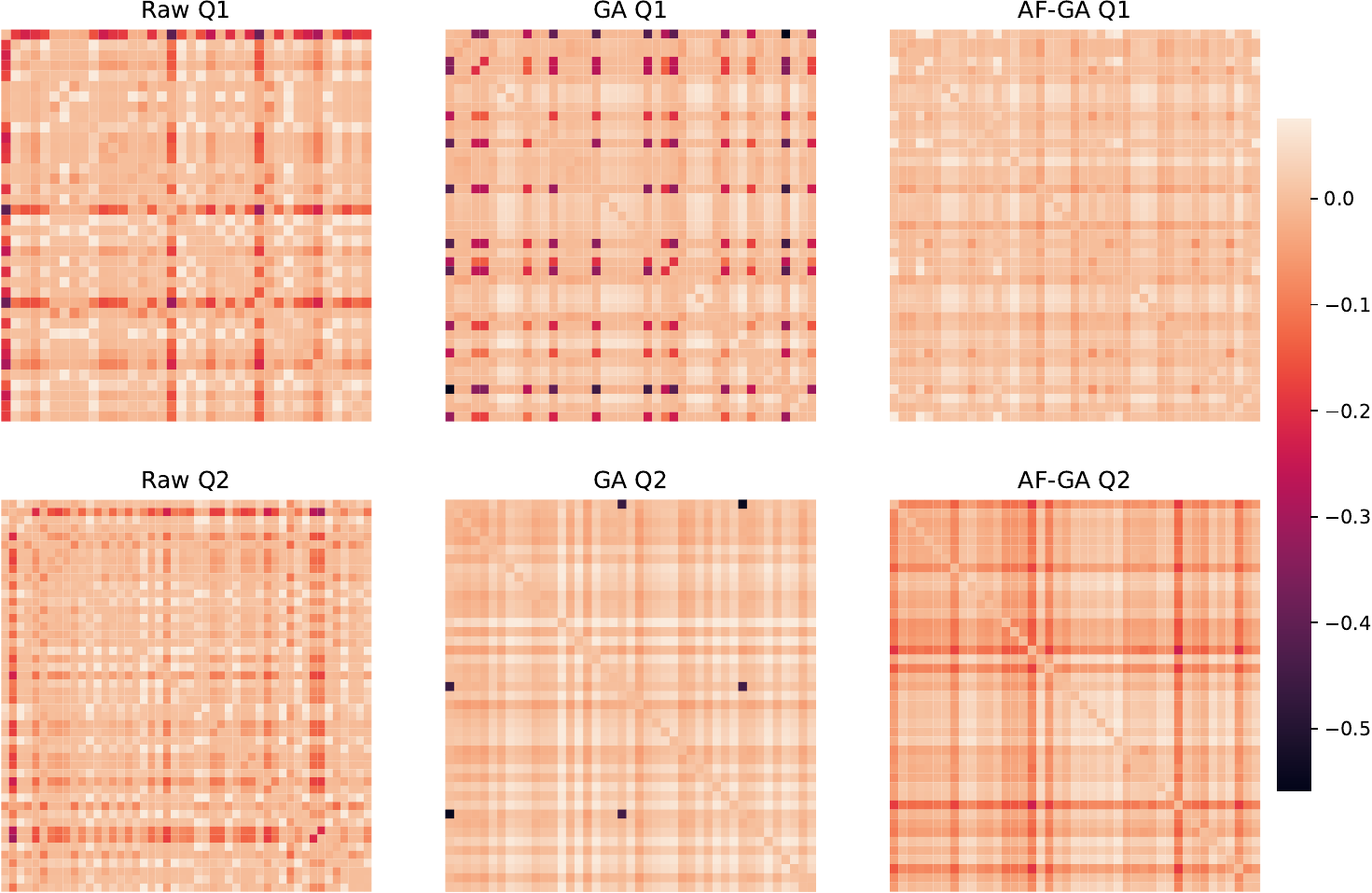}
     \caption{Classroom 45 and 46, $\phi = \rho = 1$}
     \label{grid_4546_1}
 \end{figure}

\begin{figure}[htb]
     \centering
     \includegraphics[width = 0.85\textwidth]{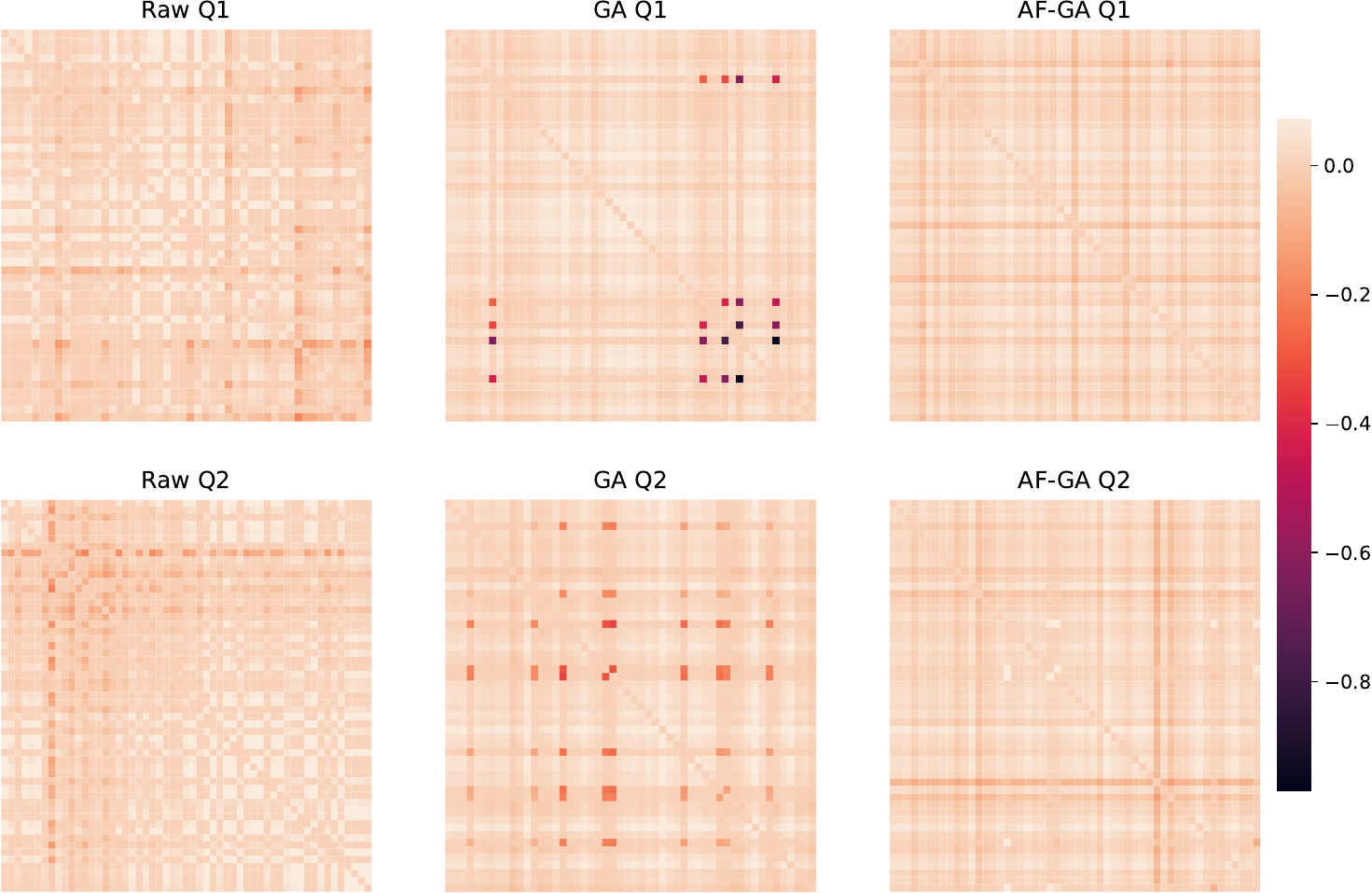}
     \caption{Classroom 69 and 70, $\phi = \rho = 0.5$}
     \label{grid_6970_0}
 \end{figure}

\begin{figure}[htb]
     \centering
     \includegraphics[width = 0.85\textwidth]{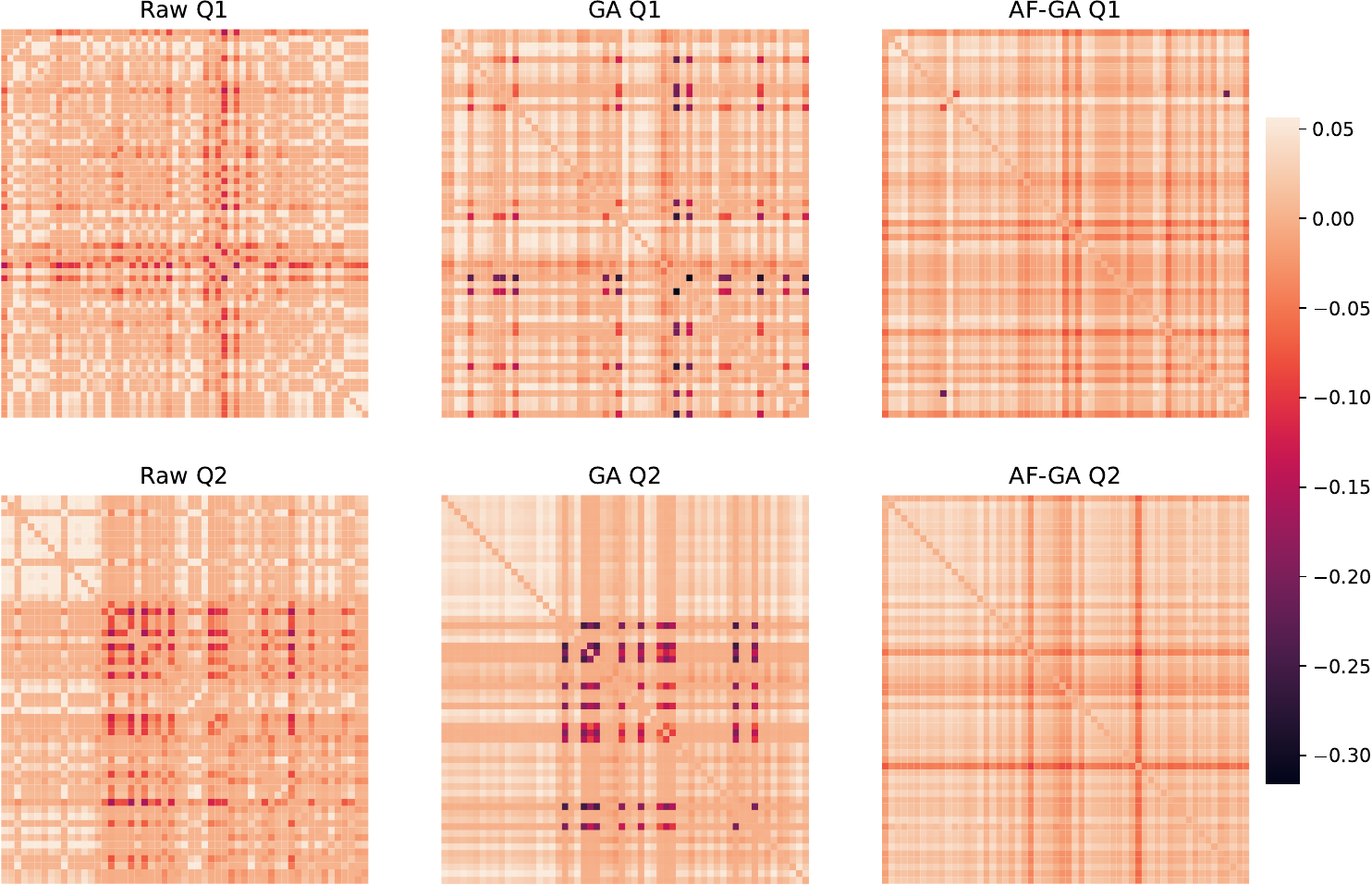}
     \caption{Classroom 73 and 74, $\phi = \rho = 0.5$}
     \label{grid_7374_0.5}
 \end{figure}

\begin{figure}[htb]
     \centering
     \includegraphics[width = 0.85\textwidth]{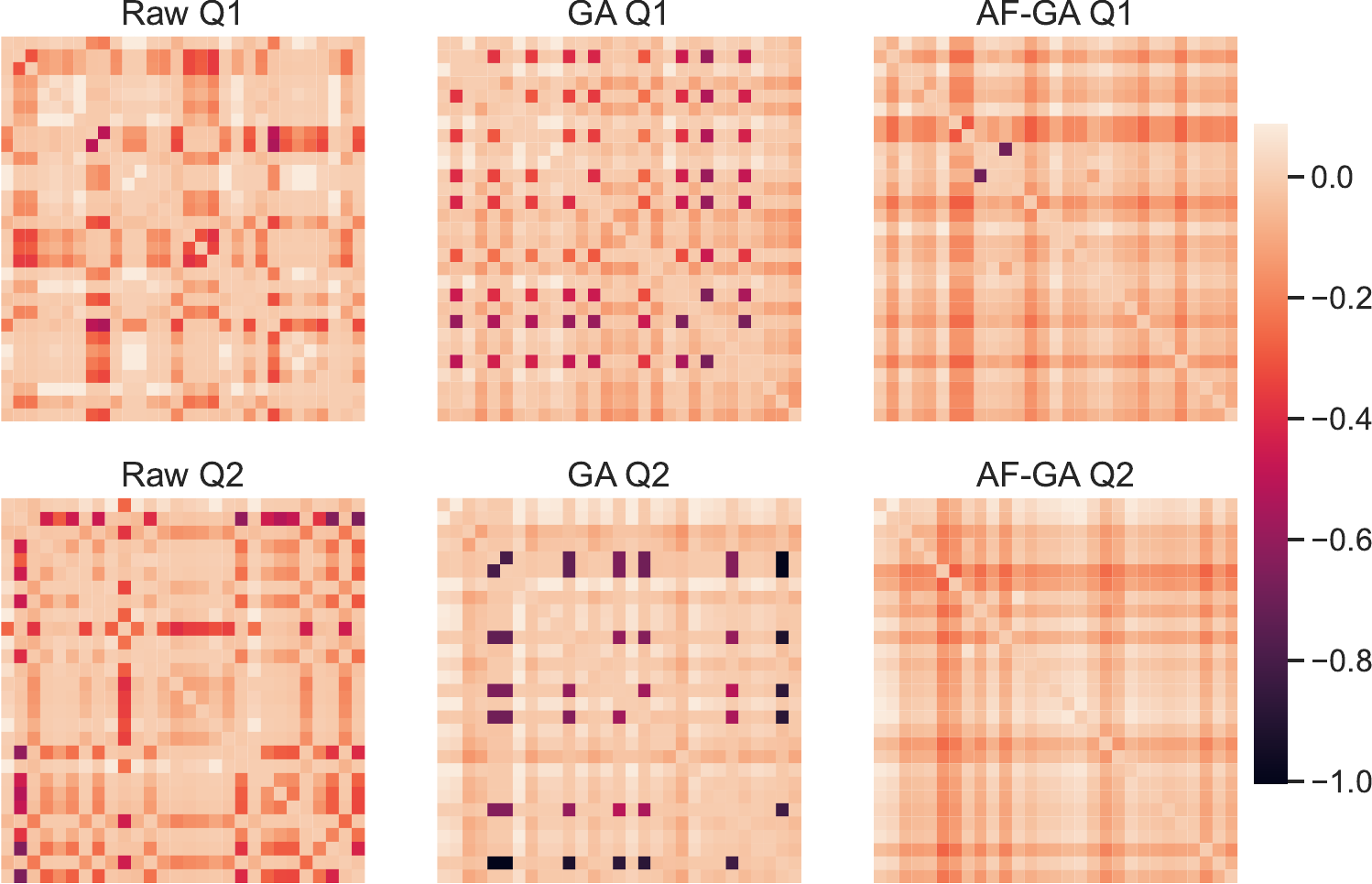}
     \caption{Classroom 77 and 78, $\phi = \rho = 1.5$}
     \label{grid_7778_1.5}
 \end{figure}

We show results for multiple schools, see Figures \ref{grid_56_1} to \ref{grid_7778_1.5}. From all figures, GA always severely penalizes a few students in pursuit of the highest average peer effect. In contrast, AFGA generates much fairer class assignment policies. In some cases, readers might be able to visually discern that the average peer effect of AFGA is higher than raw assignment, as AFGA tends to have lighter color plots.

 \setlength{\tabcolsep}{6pt}
\begin{table}[htbp]\centering
\def\sym#1{\ifmmode^{#1}\else\(^{#1}\)\fi}

\begin{tabular}{l*{6}{c}}
\toprule
            &\multicolumn{2}{c}{Gender}&\multicolumn{2}{c}{Father's education}&\multicolumn{2}{c}{Types of Hukou}\\
            &\multicolumn{1}{c}{Male} &\multicolumn{1}{c}{Female} &\multicolumn{1}{c}{Low} &\multicolumn{1}{c}{High} &\multicolumn{1}{c}{Rural} &\multicolumn{1}{c}{Others}\\
\midrule
Peer's Rank &   1.000\sym{*}&  0.946   &   1.255\sym{***}   & 0.565 & 1.215\sym{**}   & 1.143\sym{*} \\
 &     (0.581)         &     (0.592)    & (0.582)  & (0.687) &  (0.619) &  (0.650) \\
                    &&&&&& \\
Own Rank & 0.916\sym{***}  & 1.117\sym{***}  & 0.990\sym{***} &  1.037\sym{***}  & 0.984\sym{***}  & 1.008\sym{***} \\
& (0.054)  & (0.053)  & (0.053)  & (0.056) & (0.055)  &  (0.053)  \\
                   &&&&&& \\
Age      &  -0.013\sym{***}   & -0.011\sym{***}  & -0.012\sym{***} & -0.011\sym{***}  & -0.012\sym{***} & -0.012\sym{***} \\
&  (0.002)  & (0.002) & (0.002) & (0.002)  & (0.002) & (0.002) \\
                   &&&&&& \\
Sex     &    &   & 0.006  & -0.032  & 0.002 & -0.030 \\
&    &   & (0.025) & (0.024)  & (0.026)  & (0.023) \\
                   &&&&&& \\
Father's education   &  0.016\sym{*}  & 0.024\sym{***} &  &  & 0.011 & 0.019\sym{**} \\
&  (0.009)  & (0.008)  &  &   & (0.010) & (0.008) \\
                   &&&&&& \\
Mother's education    &  0.007  & 0.003  & 0.008 & 0.008  & -0.003 & 0.006 \\
& (0.009) & (0.008)  & (0.011) & (0.007)  & (0.011) & (0.008) \\
                   &&&&&& \\
Ethnic nationality   &  0.121\sym{**}  &  -0.062 & 0.033 & -0.002  & 0.100 & -0.026 \\
&  (0.062)  & (0.051)  & (0.060) & (0.052)  & (0.067) & (0.049) \\
                   &&&&&& \\
Number of observations  &   3,048    &  2,812   &  3,147     &  2,713 & 2,845  & 3,015  \\
                   &&&&&& \\
- Log Likelihood   &  3,244.771    &  2,611.928   &   3,283.262 & 2,561.519  & 2,953.243  & 2,924.823  \\
\midrule
School FE & YES & YES & YES & YES & YES & YES \\
Class RE & YES & YES & YES & YES & YES & YES \\
\bottomrule
\multicolumn{5}{l}{\footnotesize Standard errors in parentheses}\\
\multicolumn{5}{l}{\footnotesize \sym{*} \(p<0.10\), \sym{**} \(p<0.05\), \sym{***} \(p<0.01\)}\\
\end{tabular}
\caption{Heterogeneous Peer Effects of Sixth Grade Rank on Cognitive Test Score}
\label{Hetero effects full}
\end{table}

\clearpage
\section{Balance tests for class random assignment}\label{Appendix balance test}
\begin{figure}[htb]
     \centering
     \includegraphics[width = 0.65\textwidth]{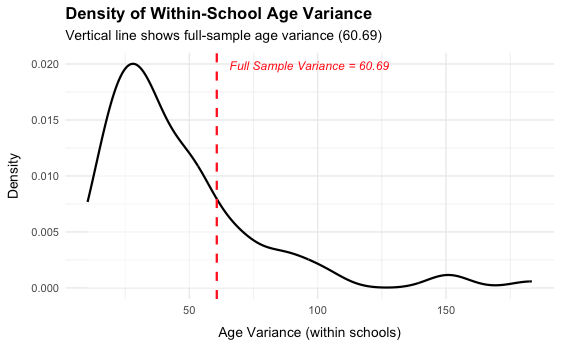}
     \caption{Density of Within-school age distribution}
     \label{density_age}
 \end{figure}

To assess whether class assignment is based on students' academic outcome, we implement a balance check regression examining the relationship between students’ own characteristics and their classmates’ 6th-grade class quantile (leave-out mean). Specifically, we regress the class quantile (excluding the individual student) on the student’s own characteristics to test whether these characteristics predict classmates’ academic ranks. The results, shown in Table \ref{balance table}, indicate that after controlling for school fixed effects, all coefficients are statistically insignificant except for Hukou type and age. The coefficient on Hukou type is marginally significant at the 5\% level. This is reasonably expected, as students within a school are typically drawn from the same geographic area and therefore tend to share the same Hukou type (rural or urban). As a result, a student’s own Hukou status may weakly predict that of their classmates, introducing a slight correlation with the classmates' class quantile. The coefficient on age is significant at the 1\% level. However, it is unlikely that schools assign students to classes based on age, particularly since grade cohorts are generally age-homogeneous. Even though age is measured in months, students within the same grade and school are expected to have similar ages. This is supported by Figure \ref{density_age}, which displays the distribution of within-school age variance. The red dashed line marks the full-sample variance of age, and the figure shows that most schools have low within-school variance, suggesting little variation in student age within schools. Therefore, these results support the assumption that class assignment within schools is random.

\begin{table}[htbp]\centering
\def\sym#1{\ifmmode^{#1}\else\(^{#1}\)\fi}
\begin{tabular}{l*{1}{c}}
\toprule
            &\multicolumn{1}{c}{Peer's Rank}\\
\midrule
Age &     -0.0005\sym{***}   \\
 &     (0.0001) \\
Peer's age & -0.022\sym{***} \\
& (0.001) \\
Sex & -0.001   \\
& (0.001) \\
Hukou & -0.002\sym{*}  \\
& (0.001)  \\
Father's education & 0.0001 \\
& (0.0003)\\
Mother's education & 0.0001  \\
& (0.0003)  \\
Ethnic nationality  & 0.0003 \\
& (0.002) \\
Number of observations  &   5,860   \\
Adjusted \(R^{2}\)&   0.192  \\
\midrule
School FE & YES \\
Class RE & NO \\
\bottomrule
\multicolumn{2}{l}{\footnotesize Standard errors in parentheses}\\
\multicolumn{2}{l}{\footnotesize \sym{*} \(p<0.10\), \sym{**} \(p<0.05\), \sym{***} \(p<0.01\)}\\
\end{tabular}
\caption{Balance test regression based on class rank}
\label{balance table}
\end{table}

\begin{figure}[htb]
     \centering
     \includegraphics[width = 1\textwidth]{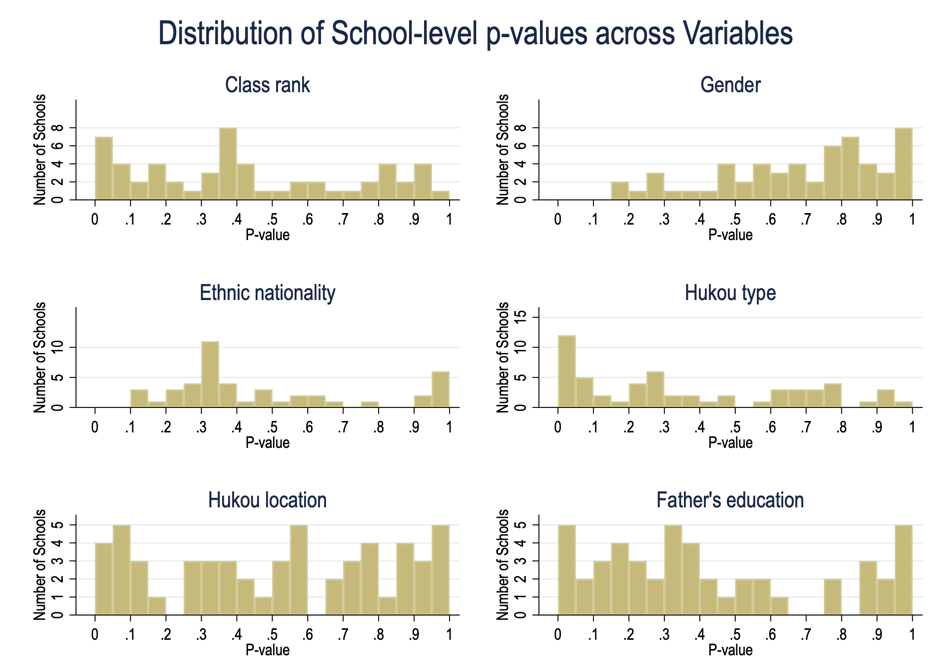}
     \caption{Distribution of P-values from Pearson $\chi^{2}$ Test}
     \label{person_chi}
 \end{figure}



\end{document}